\documentclass[twocolumn,showpacs,preprintnumbers,amsmath,amssymb]{revtex4}

\usepackage{graphicx}
\usepackage{dcolumn}
\usepackage{bm}

\def\mr#1{\mathrm{#1}}
\def\mathvec#1{\mbox{\boldmath $#1$}}

\newcommand{\pd}{\partial}  
\newcommand{\bnabla}{\pmb{\nabla}}  
\makeatletter
\@addtoreset{footnote}{page}
\makeatother
 
\begin{document}

\preprint{}

\title{High energy neutrino early afterglows \\from
gamma-ray bursts revisited}

\author{Kohta Murase}
 \email{kmurase@yukawa.kyoto-u.ac.jp}%
\affiliation{%
Yukawa Institute for Theoretical Physics, Kyoto University,\\
Oiwake-cho, Kitashirakawa, Sakyo-ku, Kyoto, 606-8502, Japan
}%

\date{\today}
                        
\begin{abstract}
The high energy neutrino emission from gamma-ray bursts (GRBs) has been
expected in various scenarios. In this paper, 
we study the neutrino emission from early afterglows of GRBs, especially 
under the reverse-forward shock model and late prompt emission model. 
In the former model, the early afterglow emission occurs due to
dissipation made by an external shock with the circumburst medium (CBM). 
In the latter model, internal dissipation such as internal shocks
produces the shallow decay emission in early afterglows. 
We also discuss implications of recent \textit{Swift} observations for
neutrino signals in detail. Future neutrino detectors such as
IceCube may detect neutrino signals from early afterglows,
especially under the late prompt emission model, while the detection would 
be difficult under the reverse-forward shock model. Contribution to
the neutrino background from the early afterglow emission may be at most 
comparable to that from the prompt emission unless the outflow making the 
early afterglow emission loads more nonthermal protons, and
it may be important in the very high energies. 
Neutrino-detections are inviting because they could provide
us with not only information on baryon acceleration but also one of the 
clues to the model of early afterglows. 
Finally, we compare various predictions for the neutrino background
from GRBs, which are testable by future neutrino-observations. 
\end{abstract}

Physics and Astronomy
\maketitle

\section{\label{sec:level1}Introduction}
Gamma-ray burst (GRB) is one of the the most energetic phenomena in
the universe and one of the candidates where both electrons and 
protons are accelerated up to very high energies. If protons are 
accelerated up to very high energies, we can expect high
energy neutrinos and gamma-rays 
that are produced by the photomeson production process.   
We investigated the high energy neutrino emission from GRBs under the
internal shock model in the previous work \cite{KM1}. Since the
prediction by Waxman \& Bahcall \cite{Wax1},
this kind of neutrino emission has been discussed by several authors
(see, e.g., \cite{Der1,Gue1,Asa1}). 

The standard (internal-external) shock model of GRBs succeeded in 
explaining many observations in the pre-\textit{Swift} era (see reviews, e.g.,
\cite{Pir1,Zha1,Pir2}). Synchrotron radiation from a  
reverse shock (RS) and a forward shock (FS) usually peaks in the
infrared-to-optical or even lower bands and ultraviolet-to-x-ray bands,
respectively. Some of the infrared/optical flashes that can be
interpreted as the RS emission were observed from GRB 
990123 \cite{Kul1,Ake1,Sar1,Mes1,Pan1,Sod1,Pan2,Fan1,Zha1a,Nak1},
GRB 021211 \cite{Fox1,Wei1,Kum1}, GRB 021004 \cite{Fox2,Kob1} and
GRB 041219a \cite{Bla1,Ves1,Fan2}. These infrared/optical photons can
interact with protons accelerated at the RS and generate high
energy neutrinos. Under the RS model, Waxman \& Bahcall predicted
high energy neutrino afterglows for the homogeneous circumburst medium
(CBM) (or interstellar medium (ISM)) \cite{Wax2}. For the wind-like 
CBM, Dai \& Lu predicted $({10}^{15} - {10}^{17})$ eV 
neutrinos \cite{Dai1}. Dermer \cite{Der2,Der1a} and Li et al. \cite{Li1} 
considered the neutrino emission from the FS, assuming that protons can be
accelerated via the second-order Fermi acceleration mechanism.
Now, large neutrino detectors such as IceCube \cite{Ahr1} and KM3Net 
\cite{Kat1} are being constructed. By exploiting these detectors, 
we can test various predictions for high energy neutrinos from GRBs 
in the near future. Furthermore, ANITA \cite{Bar1} and Auger \cite{Van1}
might also be useful for detecting very high energy neutrinos. 

Such possibilities for the neutrino emission have been studied under the 
standard model. However, the recent observations by the \textit{Swift} 
satellite have shown many unexpected behaviors in the early afterglow
phase and the simplest standard model is confronted with difficulties
 (see e.g., \cite{Gra1,Mes2,Zha2,Zha3}). 
For example, one of them is the shallow decaying behavior, which
appears after the steep decay phase. The flux-decay becomes shallow,
when the flux decays as $\propto t^{-(0.2-0.8)}$ up to a break time 
$\sim {10}^{3-4}$ s \cite{Obr1,Wil1}.
Although various modifications of the standard model have been put forward 
to explain the observations, none of which seems conclusive.

In addition, the central engine may last much longer than the duration
of the bursts. The variability of some GRB afterglows implies such 
a prolonged activity of the central engine \cite{Iok1}. 
Flares that are found in a significant fraction of \textit{Swift} 
GRBs require a long duration 
and/or re-activation of the central engine, and one of the leading 
models for explaining flares is the late internal dissipation model 
\cite{Bur1,Fal1}. Furthermore, such late internal dissipation may explain 
the shallow decaying behavior \cite{Ghi1}. After the earlier prompt
emission, there may be a tail of activity of the central engine,
producing subshells of progressively lower power and bulk
Lorentz factor for a long time.

The rapid follow-up by the \textit{Swift} satellite also allows us to observe 
many early afterglows of GRBs in the optical bands by ground-based
robotic telescopes such ROTSE. However, many ground-based and \textit{Swift} 
UVOT observations have shown that the majority of bursts have very dim
or undetectable optical afterglows \cite{Rom1}. Although foreground
 extinction, circumburst absorption and high redshift provide
explanations for many GRBs, there is tentative evidence that the strong
RS emission is suppressed. Theoretically, there are some 
possible reasons for this paucity of previously predicted optically bright
flashes, if they are ascribed to RS emission. 
One is the absence or weakness of the RS. It is because earlier
simplified estimates can over-predict the optical emission when the RS is 
semi-relativistic \cite{Nak2,Mcm1} and/or the optical emission can be much
weaker when the ejecta is highly magnetized
\cite{Zha4}. Alternatively, Beloborodov pointed out the RS emission can be 
suppressed enough if the prompt emission overlaps with the RS emission 
\cite{Bel1}. 
On the other hand, some authors suggested that the RS emission is seen as the 
shallow decay emission \cite{Uhm1,Gen1}. They argued that significant 
energy of such the RS emission may be radiated in x-ray bands and the shallow 
decay phase can be explained by the RS running into the ejecta of relatively
small Lorentz factors. 

Although many models have been proposed, there are no conclusive ones 
for explaining early afterglows. 
More and more observations will be needed in order to discriminate them.
High energy neutrino and gamma-ray observations can provide us one of 
the useful clues to such models. 
In this paper, we focus mainly on the high energy neutrino early afterglow
emission. We calculate neutrino spectra more quantitatively than
previous works under the various early afterglow models. The
method of calculation is the same as that used in our previous works 
\cite{KM2,KM3}, where we use Geant4 \cite{Ago1} with experimental data
\cite{Sch1,PDG1}. Future observations of high energy neutrinos could
provide us with not only information on baryon acceleration but also 
physical conditions of GRBs. Combined with
observations of electromagnetic emissions, they would be useful 
as one of the clues to the early afterglow models, although neutrino 
telescopes that are larger than IceCube will be desirable.

This paper is structured as follows. In Sec. \ref{subsec:levela}, we
briefly review the reverse-forward shock model. Then, we consider 
contributions from the early and late prompt emission in Sec. 
\ref{subsec:levelb} and \ref{subsec:levelc}. In Sec. \ref{sec:level2}, we 
review the high energy neutrino emission process in GRBs. 
The numerical results are shown in Sec. \ref{sec:level3}. 
We describe summary and discussions in Sec. \ref{sec:level4}.

\section{\label{sec:level2}The Model}
\subsection{\label{subsec:levela}The Reverse-Forward Shock Model}
The original reverse-forward shock model, which is successful in interpreting
many late time afterglow observations basically, has been developed 
(see reviews, e.g., \cite{Mes1,Pir1,Pir2,Zha1,Zha3}) and often accepted on
interpreting some early infrared/optical flashes such as GRB
990123. Here, we briefly review this model which is used for our
calculations \cite{Kum1,Pan2}. The analysis in this subsection is
entirely within the context of the external shock model. We use this
model only for the afterglow emission, and assume that the prompt
emission occurs due to internal dissipation such as internal
shocks. The possibility that internal dissipation makes the early
afterglow emission is described in Sec. \ref{subsec:levelc}.  
 
The expanding fireball strikes the surrounding
medium and will form two shocks: a reverse shock and a forward
shock. The shocked ambient and ejecta materials are in pressure balance
and are separated by a contact discontinuity. 
In the original standard model, the RS is
thought to be short-lived, which exists during the initial
deceleration of the fireball. After the RS crosses the ejecta, the
FS continues and the ejecta will transit to self-similar expansion 
which is described by the Blandford-Mckee solution.
 
Each shock compresses the fluid. Let the Lorentz factor of the shocked
CBM and that of the shocked ejecta (measured in the frame of
the unshocked CBM) be labeled by $\Gamma$. $\Gamma^{'}$ is the Lorentz
factor of the unshocked ejecta (measured in the frame of the shocked
ejecta), which is written as,
\begin{equation}
\Gamma ^{'} \approx \frac{1}{2} \left( \frac{\Gamma}{\Gamma _0} +
\frac{\Gamma _0}{\Gamma} \right),
\end{equation}
where $\Gamma _0$ is the Lorentz factor of the unshocked ejecta
(measured in the frame of the unshocked CBM) and we have 
assumed $\Gamma _0, \Gamma \gg 1$. By combining this equation with the 
pressure equality at the contact discontinuity, we have,
\begin{equation}
\Gamma \approx \frac{\Gamma _0}{{\left [1+2 \Gamma _0 {(n/n_{\mr{ej}})}^{1/2}
\right ]}^{1/2}}.
\end{equation}
Here the number density of the ejecta, $n_{\mr{ej}}$ is given
by $n_{\mr{ej}}=E_{\mr{ej}}/4 \pi m_{p} c^2 \Gamma _{0} (\Gamma _0
\Delta) r^2$, where $E_{\mr{ej}}$ is the isotropic energy of the ejecta
and $\Delta$ is the geometrical thickness of the ejecta measured in 
the stellar frame. The circumburst number density is written as
$n=Ar^{-k}$, where $k=0$ expresses a homogeneous CBM (ISM) and $k=2$ does
a wind-like CBM. In the latter case, $A=3 \times {10}^{35}A_{*} \, 
{\mr{cm}}^{-1}$, where $A_{*}$ is the mass-loss rate to wind speed
ratio, normalized to ${10}^{-5} M_{\odot} \, {\mr{yr}}^{-1}$. $A_{*}
\sim 1$ is a typical value for Walf-Rayet stars \cite{Che1}.

Now, we define the crossing radius $r_{\times}$, at which the RS
finishes crossing the ejecta and injection of fresh
electrons by the RS ceases. The thickness of the ejecta at the
crossing time $t_{\times}$ is written as, 
$\Delta(r _{\times})=\int _{0}^{r _{\times}} dr (\beta _{0} - 
\beta _{\mr{RS}})$, 
where $\beta _{0} - \beta _{\mr{RS}}=1.4 {\Gamma _0^2 n}^{1/2}/\Gamma _0^2
{n_{\mr{ej}}}^{1/2}$ is the difference between the speed of 
the unshocked ejecta in the stellar frame and that of 
the RS. This expression is valid for a wide range of the ratio 
$\Gamma _0^2 n/n_{\mr{ej}}$, from Newtonian to relativistic RS cases
\cite{Kum1,Pan2}. If the ejecta thickness is smaller than the 
spreading radius which is given by $r_{\mr{s}} \approx 2 \Gamma _0^2
c T$, the ejecta thickness at the crossing time 
$\Delta_{\times}$ is expressed as $\Delta_{\times} 
\approx \Delta _0 \equiv c T$. This case
is called a thick ejecta case in the usual terminology \cite{Sar2}. 
Here, $T$ is the duration of the GRB ejecta in the stellar frame. 
If the ejecta thickness is larger than $r_{\mr{s}}/ 2 \Gamma_0^2$, 
we have $\Delta_{\times} \approx r/2 \Gamma _0^2$. 
This case is called a thin ejecta case. We can write the crossing 
radius in the thick ejecta regime as follows,
\begin{equation}
r_{\times}= \left\{ \begin{array}{rl} 0.24 \times 10^{17} {\left 
( \frac{E_{\mr{ej},53}T}{n} \right)}^{1/4} \, \mr{cm} & \mbox{($k$=0)}\\
0.52 \times 10^{15} {\left ( \frac{E_{\mr{ej},53}T}{A_{*}}
\right )}^{1/2} \, \mr{cm} & \mbox{($k$=2)} \end{array} \right.
\end{equation}
In the thin ejecta regime, we have the following expressions,
\begin{equation}
r_{\times}= \left\{ \begin{array}{rl} 0.44 \times 10^{17} {\left 
(\frac{E_{\mr{ej},53}}{n \Gamma_{0,2.5}^{2}} \right )}^{1/3} \, \mr{cm} & 
\mbox{($k$=0)}\\ 
1.0 \times 10^{14} \left (\frac{E_{\mr{ej},53}}{A_{*} \Gamma _{0,2.5}^{2}}
\right ) \,\,\,\,\,\,\,\,  \mr{cm} & \mbox{($k$=2)} \end{array} \right.
\end{equation}
The essential length scales for the thick and thin ejecta cases can be
expressed as $r_{\Delta} \equiv l^{3/4}{\Delta _0}^{1/4}$ and $r_{\Gamma}
\equiv l {\Gamma _0}^{-2/3}$, respectively. Here, $l$ is the Sedov
length. For $k=0$, we have $l \equiv {(E_{\mr{ej}}/(4\pi/3)n m_p
c^2)}^{1/3}$. Using these length scales, we can write the crossing
radius $r_{\times} \sim$ max($r_{\Delta}, r_{\Gamma}$).
It is often convenient to introduce $\xi \equiv
{(r_\Gamma/r_{\Delta})}^{2} = {(l/\Delta
_0)}^{1/2}{\Gamma}^{-4/3}$. Note that, 
if $\xi \gtrsim 1$, the ejecta is essentially in the thin ejecta regime. If
$\xi \lesssim 1$, it is essentially in the thick ejecta regime.

The crossing time is calculated by 
$t_{\times}= \int_0^{r_{\times}} dr (1/2 \Gamma^2c)$.
For a thick ejecta case with $k=0$, we can obtain
$t_{\times} \approx 17 \, \mr{s} \, {\left( \frac{E_{\mr{ej},53}}{n
\Gamma_{0,2.5}^{8}} \right)}^{1/3}$.
For a thick ejecta case with $k=2$,
$t_{\times} \approx 2.9 \, \mr{s} \, \left( \frac{E_{\mr{ej},53}}{A_{\ast}
\Gamma_{0,2}^{4}} \right)$.
For thin ejecta cases with $k=0$ and $k=2$, we have $t_{\times} 
\approx 0.72 \, \mr{s} \, \times T$.
Up to an order-unity factor, the crossing time can be approximated as
$t_{\times} \sim \mr{max}[T,t_{\Gamma}]$. Here, 
the crossing time is regarded as the time when afterglows begin,
i.e., the self-similar behavior starts. 
The RS emission rises after the ejecta starts to sweep up matter,
but it does not peak until time $\sim t_{\times}$, marking the
beginning of the self-similar regime. Most of the RS emission (hence
the neutrino emission from the RS) is released at $\sim t_{\times}$. 
Therefore, we can expect that it is sufficient to consider the
behavior at $t_{\times}$ for our purpose of evaluating the neutrino energy 
fluence, although more sophisticated calculations including the time 
dependent evolution are possible \cite{Kob2}. 

The magnetic fields in each post-shock fluid are parameterized by 
fractions $\epsilon _B^f$ and $\epsilon _B^r$ 
of the post-shock internal energy density stored in it, respectively. 
They are written as,
\begin{subequations}
\begin{eqnarray}
B^f_{\times}&=&\sqrt{32 \pi \epsilon _B^f n m_p c^2 (\Gamma_{\times}
-1)(\Gamma_{\times}+3/4)} \\
B^r_{\times}&=&\sqrt{32 \pi \epsilon _B^r n_{\mr{ej}} m_p c^2 
(\Gamma ^{'}_{\times}-1)(\Gamma ^{'}_{\times}+3/4)}
\end{eqnarray}
\end{subequations}
The injection Lorentz factor of electrons at the crossing time is,
\begin{equation} 
\gamma _{e,m}= \frac{\epsilon _e}{f_e} g(p) \frac{m _p}{m_e} 
(\Gamma ^{(')}_{\times} -1),
\end{equation}
where $g(p)$ is given by $g(p)=(p-2)/(p-1)$ and $f_e$ is a number fraction of
the shocked electrons that are injected into the acceleration process. 
The cooling Lorentz factor of electrons is obtained by equating 
the cooling time and dynamical time 
$t_{\mr{dyn}}=r_{\times}/\Gamma_{\times}c$, which is written as,
\begin{equation}
\gamma _{e,c} = \frac{6\pi m_e c^2 \Gamma_{\times}}
{\sigma _{\mr{T}} B_{\times}^2 r_{\times} (1+Y)},
\end{equation}
where $Y$ is the Compton parameter, which can be calculated from the
electron distribution (see e.g., \cite{Pan4}).

The injection energy $\varepsilon^m = h \nu_{m}$ and cooling energy
$\varepsilon^c = h \nu_{c}$ in the comoving frame are calculated from the
corresponding electron Lorentz factors by using $\varepsilon = h \nu \approx
\hbar \gamma _e^2\frac{eB_{\times}}{m_e c}$.
In this paper, we consider the four cases; the thick ejecta colliding into
the ISM, the thin ejecta colliding into the ISM, 
the thick ejecta colliding into the wind-like CBM, and 
the thin ejecta colliding into the wind-like CBM. 
For example, let us consider the case of the thin ejecta colliding
into the ISM. We can obtain two characteristic frequencies 
in the observer frame ($\nu^{\mr{ob}} \simeq \Gamma_{\times} \nu$) as follows. 
For the FS emission, we have 
$\nu_{m,f}^{\mr{ob}} \approx 2.9 \times {10}^{20} \, \mr{Hz} 
\, g^2 {({\epsilon}_{B,-2}^{f})}^{1/2} 
  ({\epsilon_{e,-1}^{f})}^2 {(f_{e}^f)}^{-2} E_{\mr{ej},53}^{1/2}
T^{-1/2}$ and $\nu_{c,f}^{\mr{ob}} \approx 4.5 \times {10}^{16} \, \mr{Hz} 
\, {(\epsilon_{B,-2}^{f})}^{-3/2} n_0^{-1} E_{\mr{ej},53}^{-1/2} T^{-1/2} 
{(1+Y)}^{-2}$. For the RS emission, we have $\nu_{m,r}^{\mr{ob}} 
\approx 1.0 \times {10}^{18} \, \mr{Hz} \, g^2 {({\epsilon}_{B}^r)}^{1/2} 
  {(\epsilon_{e}^r)}^2 {(f_{e}^r)}^{-2}
{(\Gamma_{\times}^{\prime}-1)}^2 n_0^{3/8} E_{\mr{ej},53}^{3/8} T^{-1/8}$, 
$\nu_{c,r}^{\mr{ob}} \approx 4.5 \times {10}^{13} \, \mr{Hz} \, 
{(\epsilon_{B}^r)}^{-3/2} n_0^{-1} E_{\mr{ej},53}^{-1/2} T^{-1/2} 
{(1+Y)}^{-2}$.
Basically, the observed cooling frequencies of RS and FS emissions are
equal if microphysical parameters are similar, while the observed injection
frequency of the RS emission is smaller than that of the FS emission by a factor of 
$\Gamma^2_{\times}$. 

We also consider the synchrotron self-absorption process.
The optical thickness for synchrotron self-absorption can be
approximated by (see, e.g., \cite{Pan4}),
\begin{equation}
\tau _{sa}(\varepsilon) \simeq \frac{5 e }{ 4 \pi r_{\times}^2 B_{\times} 
\gamma _{e,n}^5} f_e N_e \times 
\left\{ \begin{array}{rl}{(\varepsilon/\varepsilon^{n})}^{-\frac{5}{3}}
\,\,\,\,\,\,\,\,\, & \mbox{($\varepsilon < \varepsilon^n$)}\\
{(\varepsilon/\varepsilon^{n})}^{-\frac{(p+4)}{2}} 
& \mbox{($\varepsilon ^n \leq \varepsilon$)} \end{array} \right.
\end{equation}
where 
\begin{subequations}
\begin{eqnarray}
N_e^f &=& \frac{4\pi}{3-k} A r^{3-k} \\
N_e^r &=& \frac{E_{\mr{ej}}}{\Gamma _0 m_p c^2}
\end{eqnarray}
\end{subequations}
are the number of electrons energized by the FS and RS, respectively. 
$\gamma _{e,n} \equiv \mr{min}(\gamma _{e,c}, \gamma _{e,m})$, 
$\varepsilon^n \equiv \mr{min}(\varepsilon^m, \varepsilon^c)$, and we have
assumed $\gamma_{e,n} > 1$. 
The self-absorption energy is determined by
$\tau_{sa}(\varepsilon^{sa})=1$. Hence, we can obtain,
\begin{equation}
\varepsilon^{sa} \simeq \varepsilon^n 
\left\{ \begin{array}{rl}{\tau_{sa}}^{\frac{3}{5}}
(\varepsilon=\varepsilon^{n})
\,\,\,\,\,\,\,\,\, & \mbox{($\varepsilon < \varepsilon^n$)}\\
{\tau_{sa}}^{\frac{2}{(p+4)}}(\varepsilon=\varepsilon^{n}) & 
\mbox{($\varepsilon ^n \leq \varepsilon$)} \end{array} \right.
\end{equation}
Generally, the observed self-absorption frequency in the RS is typically 
higher than that in the FS.

The comoving specific luminosity per unit energy at the injection
energy or cooling energy is approximated by 
\begin{equation}
L_{\varepsilon, \mr{max}}= \frac{1}{2 \pi \hbar} f_e N_e \frac{\sqrt{3} e^3
B}{m_e c^2} \phi_p,
\end{equation}
where $\phi _p$ is an order-unity coefficient calculated by Wijers \&
Galama \cite{Wij1}. In this paper we use $\phi_p \approx 0.6$.
For example, let us consider the case of the thick ejecta 
colliding into the ISM. The observed peak luminosity per unit energy
from the FS emission is written as $L_{\varepsilon_{\mr{ob}},\mr{max}}^f =
 1.3 \times {10}^{58} \, {\mr{s}}^{-1} \, f_e^f E_{\mr{ej},53} 
 {({\epsilon_{B,-2}^f})}^{1/2} n_{0}^{1/2} T^{1/2} (\phi_p/0.6)$, and 
that from the RS emission is expressed as 
$L_{\varepsilon_{\mr{ob}},\mr{max}}^r = 4.9 \times {10}^{62} \,
\mr{s}^{-1} \, f_e^r E_{\mr{ej},53}^{5/4} {({\epsilon_B^r})}^{1/2}
n_{0}^{1/4} \Gamma_{0,2.5}^{-1} T^{-1/4} (\phi_p/0.6)$.
The observed peak flux from the RS is typically larger by a factor
of $\Gamma_{\times}$ than that from the FS.

From the above equations, we can obtain photon spectra. 
We have for $\varepsilon^{sa}< \varepsilon^m
<\varepsilon^c$ (the slow cooling regime),
\begin{equation}
\frac{dn}{d\varepsilon} = n_{\varepsilon,\mr{max}} 
\left\{ \begin{array}{ll}
{(\varepsilon^{sa}/\varepsilon ^{m})}^{-\frac{2}{3}} 
{(\varepsilon/\varepsilon ^{sa})}^{1} & 
\mbox{( $\varepsilon ^{\mr{min}} \leqq \varepsilon \leqq \varepsilon ^{sa}$)}\\
{(\varepsilon/\varepsilon ^{m})}^{-\frac{2}{3}} & 
\mbox{($\varepsilon ^{sa} < \varepsilon \leqq \varepsilon ^m$)}\\
{(\varepsilon/\varepsilon^{m})}^{-\frac{p+1}{2}} & 
\mbox{($\varepsilon ^ m< \varepsilon \leqq \varepsilon ^{\mr{c}}$)}\\
{(\varepsilon^c/\varepsilon^m)}^{-\frac{p+1}{2}} 
{(\varepsilon/\varepsilon^c)}^{-\frac{p+2}{2}} & \mbox{($\varepsilon ^c
< \varepsilon \leqq \varepsilon ^{\mr{max}}$)},
\end{array} \right.
\end{equation}
and for $\varepsilon^{sa}< \varepsilon^c <\varepsilon^m$ (the fast 
cooling regime), 
\begin{equation}
\frac{dn}{d\varepsilon} = n_{\varepsilon,\mr{max}} 
\left\{ \begin{array}{ll}
{(\varepsilon^{sa}/\varepsilon ^{c})}^{-\frac{2}{3}} 
{(\varepsilon/\varepsilon ^{sa})}^{1} & 
\mbox{($\varepsilon ^{\mr{min}} \leqq \varepsilon \leqq \varepsilon ^{sa}$)}\\
{(\varepsilon/\varepsilon ^{c})}^{-\frac{2}{3}} & 
\mbox{($\varepsilon ^{sa} < \varepsilon \leqq \varepsilon ^{c}$)}\\
{(\varepsilon/\varepsilon^{c})}^{-\frac{3}{2}} & 
\mbox{($\varepsilon ^{c}< \varepsilon \leqq \varepsilon ^{m}$)}\\
{(\varepsilon^m/\varepsilon^c)}^{-\frac{3}{2}} 
{(\varepsilon/\varepsilon^m)}^{-\frac{p+2}{2}} & 
\mbox{($\varepsilon ^m < \varepsilon \leqq \varepsilon ^{\mr{max}}$)}.
\end{array} \right.
\end{equation}
Similarly we can obtain for 
$\varepsilon^{m}< \varepsilon^{sa} <\varepsilon^c$, 
\begin{equation}
\frac{dn}{d\varepsilon} = n_{\varepsilon,\mr{max}} 
\left\{ \begin{array}{ll}
{(\varepsilon^{sa}/\varepsilon^{m})}^{-\frac{p+1}{2}}
{(\varepsilon^{m}/\varepsilon ^{sa})}^{\frac{3}{2}} 
{(\varepsilon/\varepsilon ^{m})}^{1} 
\\
{(\varepsilon^{sa}/\varepsilon^{m})}^{-\frac{p+1}{2}}
{(\varepsilon/\varepsilon ^{sa})}^{\frac{3}{2}} 
\\
{(\varepsilon/\varepsilon^{m})}^{-\frac{p+1}{2}} 
\\
{(\varepsilon^c/\varepsilon^{m})}^{-\frac{p+1}{2}}
{(\varepsilon/\varepsilon^c)}^{-\frac{p+2}{2}} 
,
\end{array} \right.
\end{equation}
and for $\varepsilon^{c}< \varepsilon^{sa} <\varepsilon^m$ (where we
do not consider the inhomogeneity \cite{Gra2}), 
\begin{equation}
\frac{dn}{d\varepsilon} = n_{\varepsilon,\mr{max}} 
\left\{ \begin{array}{ll}
{(\varepsilon^{sa}/\varepsilon^{c})}^{-\frac{3}{2}}
{(\varepsilon^{c}/\varepsilon ^{sa})}^{\frac{3}{2}} 
{(\varepsilon/\varepsilon ^{c})}^{1} 
\\
{(\varepsilon^{sa}/\varepsilon^{c})}^{-\frac{3}{2}}
{(\varepsilon/\varepsilon ^{sa})}^{\frac{3}{2}} 
\\
{(\varepsilon/\varepsilon^{c})}^{-\frac{3}{2}} 
\\
{(\varepsilon^{m}/\varepsilon^{c})}^{-\frac{3}{2}}
{(\varepsilon/\varepsilon^m)}^{-\frac{p+2}{2}} 
,
\end{array} \right.
\end{equation}
where
\begin{equation}
n_{\varepsilon, \mr{max}}= \frac{L_{\varepsilon,\mr{max}}}{4 \pi
 r_{\times}^2 \Gamma_{\times}^2 c {\varepsilon}^{n}}.
\end{equation}
In order to demonstrate neutrino spectra from the RS, we adopt
following parameter sets.

ISM-tc: $\Gamma_{0}={10}^{2.5}$, $E_{\mr{ej}}=4 \times {10}^{53}$
ergs, $\Delta_0=4.5 \times{10}^{11}$ cm, $n=5 \, \mr{cm}^{-3}$, 
 $\epsilon_{B}^r=0.01$, $\epsilon_e^r=1/4$, $f_e^r=1$ and $p=2.4$. 
This is one of the parameter sets representing the thick ejecta that
collides into the ISM.
 
ISM-tn: $\Gamma_{0}={10}^{2}$, $E_{\mr{ej}}=4 \times {10}^{52}$
ergs, $\Delta_0=4.5 \times{10}^{11}$ cm, $n=0.5 \, \mr{cm}^{-3}$, 
 $\epsilon_{B}^r=0.01$, $\epsilon_e^r=1/4$, $f_e^r=1$ and $p=2.4$. 
This is one of the parameter sets representing the thin ejecta that
collides into the ISM.

ISM-e: $\Gamma_{0}=300$, $E_{\mr{ej}}={10}^{55}$
ergs, $\Delta_0=1.05 \times{10}^{12}$ cm, $n=1 \, \mr{cm}^{-3}$, 
 $\epsilon_{B}^r=0.001$, $\epsilon_e^r=0.04$, $f_e^r=1$ and $p=2.5$. 
This is one of the parameter sets representing the thin ejecta that
collides into the ISM \cite{Pan2}. 

ISM-eb: $\Gamma_{0}=300$, $E_{\mr{ej}}={10}^{54}$
ergs, $\Delta_0=1.05 \times{10}^{12}$ cm, $n=1 \, \mr{cm}^{-3}$, 
 $\epsilon_{B}^r=0.2$, $\epsilon_e^r=0.1$, $f_e^r=1$ and $p=2.5$. 
This is one of the parameter sets representing the thick ejecta that
collides into the ISM \cite{Zha1a}. 
Note that the models ISM-e and ISM-eb express cases where
the ejecta is energetic $E_{\rm{ej}} \gtrsim {10}^{54}$ ergs. GRB
990123 is thought as such an energetic event. (Note that, if the ejecta
carries the magnetic field directly from the central engine,
$\epsilon_B^r$ can be significantly larger than $\epsilon_B^f$. A strong reverse
shock is expected when the magnetic field is radiationally important
but not yet dynamically important. GRB 990123 is thought to be one of
such events \cite{Zha1a}.)

WIND-tc: $\Gamma_{0}={10}^{2.5}$, $E_{\mr{ej}}=4 \times {10}^{53}$
ergs, $\Delta_0=4.5 \times{10}^{11}$ cm, $A_{\ast}=1$, 
 $\epsilon_{B}^r=0.01$, $\epsilon_e^r=1/4$, $f_e^r=1$ and $p=2.4$. 
This is one of the parameter sets representing the thick ejecta that
collides into the wind-like CBM. 

WIND-tn: $\Gamma_{0}={10}^{2}$, $E_{\mr{ej}}=4 \times {10}^{52}$
ergs, $\Delta_0=4.5 \times{10}^{11}$ cm, $A_{\ast}=0.01$, 
 $\epsilon_{B}^r=0.01$, $\epsilon_e^r=1/4$ and $p=2.4$. This is one of
the parameter sets representing the thin ejecta that collides into
the wind-like CBM.

So far, we have assumed $f_e=1$. However, this might not be 
true \cite{Eic1}. We can consider cases with $f_e \ll 1$. If a small
fraction of the electron population can have a significant part of 
the dissipated energy in the RS, the RS emission can appear in x-rays 
\cite{Gen1}. 
The RS emission, which has usually been used for explanation of
infrared/optical flashes that are observed for only a small fraction of
GRBs, might play an important role in the early afterglow phase. 
Recently, some authors proposed that the 
mysterious shallow decay emission can be explained by the RS emission 
\cite{Gen1,Uhm1}.
In their models, the plateau shape can be achieved by requiring the
appropriate distribution of Lorentz factors of the ejecta. It is
assumed that Lorentz factors of the material, which is ejected 
during the last stages of source activity, decrease to small values 
of Lorentz factors. The head of the ejecta has larger Lorentz factors, 
while the tail of the ejecta has smaller Lorentz factors. In such cases, 
a long-lived RS is possible when it propagates into the stratified
ejecta with decreasing Lorentz factors. In addition, this model also
requires that the forward shock emission can be negligible at least in
the early afterglow phase. The suppression of the FS emission might 
occur because the magnetic fields are too weak in the
external medium and not sufficiently amplified \cite{Mil1}, and/or the 
first-order Fermi acceleration mechanism is not so efficient.
To reproduce the shallow decaying behavior, 
we need the detailed numerical modeling. This is beyond scope of this
paper because our goal is not to explain the shallow decay emission.  
For our purpose to estimate neutrino fluxes, it will be sufficient
to consider the RS emission by the head of the ejecta. Assuming that
the head of the ejecta carries the energy $E_{\mr{ej}}^h = (1/3)
E_{\mr{ej}}$, we adopt the following parameter sets.  
   
ISM-s: $\Gamma_{0}={10}^{2.5}$, $E_{\mr{ej}}^h = \frac{4}{3} \times {10}^{53}$
ergs, $\Delta_0=4.5 \times{10}^{11}$ cm, $n=5 \, \mr{cm}^{-3}$, 
 $\epsilon_{B}^r=1/4$, $\epsilon_e^r=1/4$, $f_e^r=0.025$ and $p=2.4$. 
This is one of the parameter sets representing the thick ejecta that
collides into the ISM.

WIND-s: $\Gamma_{0}={10}^{2.5}$, $E_{\mr{ej}}^h= \frac{4}{3} \times {10}^{53}$
ergs, $\Delta_0=4.5 \times{10}^{11}$ cm, $A_{\ast}=0.1$, 
 $\epsilon_{B}^r=1/4$, $\epsilon_e^r=1/4$, $f_e^r=0.025$ and $p=2.4$. 
This is one of the parameter sets representing the thick ejecta that
collides into the wind-like CBM. 

In order to predict neutrino fluxes, we have to set the amount of accelerated
protons which no one knows from the first principle. In this paper, we 
just assume that the moderately efficient acceleration occurs and take 
$\epsilon_{\mr{acc}}=1/4$, where $\epsilon_{\mr{acc}} 
\equiv \zeta_p(1-\epsilon_{B}-\epsilon_{e})$ and $\zeta_{p}$ is  
acceleration efficiency. 

\subsection{\label{subsec:levelb}The Overlapping of Prompt Emission with
the Shocked Region}
In the previous subsection, we have considered the RS emission under the
original RS model, which typically predicts infrared/optical flashes. 
As previously noted, there is tentative evidence of the lack of 
infrared/optical flashes \cite{Rom1}. Several possible reasons 
have been suggested.
First, the ejecta may be strongly magnetized \cite{Zha4}. Then, the
hydrodynamical shock can become weak or there is no RS. Second, especially 
in the thin ejecta, the RS emission can be more suppressed than 
the earlier simplest estimations \cite{Nak2,Mcm1}. Such a semi-relativistic RS 
may give a peak flux below one give by a FS. In addition, 
a pair-rich RS may be common \cite{Mcm1}. Third, in the thick ejecta, 
The RS emission can be suppressed because RS electrons are rapidly 
cooled due to Compton scattering by photons from the prompt emission \cite{Bel1}.

Now, we consider the third possibility, i.e., overlapping of the
prompt emission with the shocked region. This overlapping will occur
for the thick ejecta case, which can be expected for long GRBs we
consider throughout this paper. The prompt emission which occurs due
to internal dissipation can provide additional target photons for 
accelerated protons, so that more neutrinos can be produced via the 
photomeson production process. This possibility 
was suggested by Fan et al. \cite{Fan3} for the thick ejecta colliding
into the wind-like CBM. Here, we also study such possibilities
including cases of the ejecta that collides with ISM in more detail. 
We assume $E_{\gamma}^{\mr{iso}} = 10^{53}$ ergs as the
isotropic prompt emission energy. 
The averaged photon energy density within the ejecta at the crossing
radius is given by,
\begin{equation}
U_{\gamma} \approx \frac{ {\Gamma^{\prime}}^2 
E_{\gamma}^{\mr{iso}}}{4\pi r_{\times}^2 \Delta_0 \Gamma _0^2}.
\end{equation}
A spectrum of the prompt emission is well approximated by a broken 
power-law spectrum, which is,
\begin{equation}
\frac{dn}{d\varepsilon} \propto \left\{ \begin{array}{rl} 
{\varepsilon}^{-\alpha} & \mbox{(for $\varepsilon ^{\mr{min}} \leq \varepsilon 
< \varepsilon ^b$)}\\  
{\varepsilon}^{-\beta} & \mbox{(for $\varepsilon ^b \leq \varepsilon 
\leq \varepsilon ^{\mr{max}}$)}. 
\end{array} \right.
\end{equation}
The observed break energy is $\sim 250 \, \mr{keV}$, which corresponds
to the break energy in the comoving frame, $\varepsilon ^{b} \sim$ a
few $\mr{keV}$. Hence, we set $\varepsilon ^{b} =1 \, \mr{keV}$ and 
spectral indices to $\alpha = 1$, $\beta = 2.2$ similarly to our
previous work \cite{KM1}. We take the minimum energy as $1 \, \mr{eV}$ and 
the maximum energy as $10 \, \mr{MeV}$ in the comoving frame.

If such a photon flow due to the prompt emission cools down high energy
electrons sufficiently, strong infrared/optical flashes
that are expected in the RS model can be suppressed \cite{Bel1}.
In the model ISM-tc which represents the thick ejecta case,
 the cooling Lorentz factor of electrons $\gamma_{e,c}$ is
significantly lowered due to Compton scattering by prompt
photons. It leads to that the RS emission is not in the slow cooling
regime but in the fast cooling regime.
 
In this paper, we consider the four parameter sets; ISM-tc2, ISM-s2, 
WIND-tc2 and WIND-s2. Model parameters are the same as those in
ISM-tc, ISM-s, WIND-tc and WIND-s, respectively. But the overlapping
effect due to the prompt emission is included. A target photon spectrum is 
given by a superposition of a prompt spectrum and a RS spectrum
modified by the overlapping effect.
Note that the high energy gamma-ray emission can be expected in this
model \cite{Bel1,Fan3}. These up-scattered photons can also contribute to
the photomeson production process, but we can neglect such a population
because of a smaller number of these high energy photons. 

\subsection{\label{subsec:levelc}The Late Prompt Emission Model}
The late internal dissipation may last longer than the duration of
earlier internal dissipation that makes the prompt emission. For example, 
some of the flares are likely to be attributed to the late internal 
dissipations \cite{Bur1,Fal1}. 
The flares typically happen hundreds of seconds
after the trigger or earlier. In some cases, they occur around a
day after the main burst. The amplitudes of the flares are usually larger than the underlying
afterglow component by a factor of several, but can be much
larger. Recent analyses show that the averaged radiation energy of flares is 
approximately $\sim 1/10$ of that of the prompt emission \cite{Fal1}.  

One of the leading models for such late internal dissipation 
is the late internal shock model (see, e.g., \cite{Fan4}).   
Lorentz factors of ejected subshells will be highly
variable. If $\Gamma _{\mr{s}} \sim 10$ and $\Gamma _{\mr{f}}
\sim 100$ are typical Lorentz factors of the slow and fast subshells 
respectively, the Lorentz factor of the merged subshell can be
expressed as $\Gamma _{0} \approx \sqrt{\Gamma _{\mr{f}} 
\Gamma_{\mr{s}}} \sim 30$. The Lorentz factor of the internal shocks
can be estimated by $\Gamma _\mr{sh} \approx 
(\Gamma _f/\Gamma _s + \Gamma _s/\Gamma _f)/2 \sim$ a few.
The typical collision radius is given by,
$r \approx 2\Gamma _{0}^2 c \delta t \approx {10}^{15.3}  \, \mr{cm}
\, {(\Gamma _{0}/15)}^{2}[\delta t/150(1+z) \, \mr{s}]$.  
The internal shocks are expected to be mildly relativistic shocks. If
protons are accelerated efficiently in these shocks, neutrinos can
appear through the photomeson production process. In our previous work
\cite{KM2}, we predicted such high energy neutrino flashes from flares
of GRBs.

On the other hand, the x-ray emission in the shallow decay phase is 
often attributed to the external shock emission. 
However, Ghisellini et al. \cite{Ghi1} recently suggested 
that this plateau phase for the x-ray emission may be due to the late 
prompt emission. In their model, this late prompt emission can be due
to the same internal dissipation process as that for the early prompt 
emission but by subshells created at late times
with smaller $\Gamma_0$ and much lower power. The radiation can be
produced at distances relatively close to the central engine (even
less than $r \sim {10}^{13-14}$ cm), in a different region where the
subshells interact with the CBM \cite{Ghi1}.

Although this model can explain the chromatic behavior in early
afterglows, it has not enabled us to explain the closure relations, which are
expected in forward shock models and satisfied for the normal decay segment
following the shallow decay segment. In spite of such a defect, there 
are indeed a couple of bursts that are likely to show the 
late prompt emission, marked by a sharp decay following an extended
plateau with flickering. Such a striking behavior is seen in some
bursts such as GRB 070110 \cite{Tro1a}. 
   
If the shallow decay emission is attributed to the late prompt emission and 
late internal dissipation is due to late internal shocks or
other models that allow a significant fraction of baryons to be
accelerated up to sufficiently high energies, we can also expect high 
energy neutrino signals coincident with the shallow decay emission like the case of
flares. 
The neutrino emission
due to late internal dissipation itself would decay during the shallow and
normal decay phases after the late prompt emission starts. 
Note that, physical conditions may be similar to those expected
in flares and flares may be produced by a late shell, moving with a
somewhat larger Lorentz factor than the shells created just earlier 
\cite{Ghi1}. Therefore, we can use the same framework as that used in 
Murase \& Nagataki \cite{KM2} in order to evaluate high energy neutrino 
fluxes associated with the early x-ray emission in the shallow decay
phase under the late prompt emission model. In this paper,
let us refer neutrinos from both flares and the late 
prompt emission as neutrinos from the late prompt emission.  

The photon energy density is given by,
\begin{equation}
U_{\gamma}=\frac{E_{\gamma,\mr{sh}}^{\mr{iso}}}{4\pi \Gamma _{0} r^2 l}
\end{equation}
where $l$ is the width of subshells and $E_{\gamma,\mr{sh}}^{\mr{iso}}$ is the
radiated energy from each subshell. $l$ is typically given by
$l=r/\Gamma_0$, although it can be a smaller value.
We assume the relatively small Lorentz factor, $\Gamma_0 \sim$ a few$\times 10$.
The emitted energy in the shallow decay phase is typically $\sim 1/10$
of that of the prompt emission \cite{Lia1}. 
Hence, we adopt the total isotropic radiation
energy of the x-ray emission in the shallow decay phase, 
$E_{\mr{LP}}^{\mr{iso}} \equiv N E_{\gamma, \mr{sh}}^{\mr{iso}}= 
{10}^{52}$ ergs, where $N$ is a number of collisions between ejected subshells.
The shallow decay emission typically occur at $T \sim {10}^{3-4}$ s, 
and a collision radius can be  $r \sim {10}^{13-16}$ cm under the late
prompt emission model (and note that the expected variability time scale
under the late internal shock model can be estimated by using 
$\delta t \approx (1+z)r/2\Gamma_0^2 c$). Hence, we take 
$N \sim$ (a few$-100$).

For numerical calculations, we assume a broken power-law spectrum similarly
to the cases of the prompt emission, and we use Eq. (18). But we take $\varepsilon
^b = (10-100)$ eV. We also set ${\varepsilon}^{\mr{min}} =0.1$ eV and 
${\varepsilon}^{\mr{max}}=1$ MeV.

Magnetic energy density is expressed as $U_B \equiv \xi_B
U_{\gamma}$  where $\xi_B \approx \epsilon_B/\epsilon_e$. 
In this paper, we set $\xi_B=1$.
Nonthermal proton energy density is parameterized as $U_{p} \equiv 
\xi_{\mr{acc}}U_{\gamma}$, 
where $\xi_{\mr{acc}} \approx \epsilon_{\mr{acc}}/\epsilon_e =
\zeta_p(1-\epsilon_B-\epsilon_e)/\epsilon_e$ is a nonthermal baryon loading
factor, where $\zeta_p$ is the proton acceleration efficiency. 
The present acceleration theory cannot give this value from the 
first principle. Here, we assume that protons can be accelerated
efficiently as is usually expected in supernova remnants and just 
adopt $\xi_{\mr{acc}}=10$ as a fiducial value. 
If the proton distribution has $d n_p / d \varepsilon_p \propto 
\varepsilon_p^{-2}$ and the minimum energy of protons is a few$\times
m_p c^2$, this fiducial value $\xi_{\rm{acc}}=10$ means that 
the energy density of nonthermal protons with energy
$\varepsilon_{p}$, $\varepsilon_p^2 {(\frac{d n_p}{ 
d \varepsilon_p})}_{\varepsilon_p}$ is comparable to 
the radiation energy density of photons with energy $\varepsilon^b$, 
${(\varepsilon^b)}^2 {(\frac{d n }{d
\varepsilon})}_{\varepsilon=\varepsilon^b}$. A similar
assumption is often adopted in previous works \cite{KM1,Wax1,Asa1,Gue1}.
If the nonthermal baryon loading factor could be larger, we expect
higher neutrino fluxes, but no one obtains this value from the first 
principle. On the other hand, there is a hypothesis that UHECRs come from
GRBs based on the internal shock model that will cause the usual prompt
emission \cite{Wax1}, which typically requires sufficiently large
nonthermal baryon loading factors $\xi_{\rm{acc}} \sim (50-100)$
\cite{KM1} (although statements depend on the evaluation of the local
GRB rate which has some uncertainties \cite{Gue2}). Motivated by this 
hypothesis, we also use $\xi_{\mr{acc}}=50$ as the optimistic value 
in this paper. 

We adopt the following parameter sets to estimate the neutrino flux
from GRBs in the late prompt emission model. 

LP0: $E_{\gamma,\mr{sh}}^{\mr{iso}} = 10 ^{51.2} \, \mr{ergs}$, $\Gamma_0=15$
$r={10}^{15.3} \, \mr{cm}$, $\xi _B =1$ and $\varepsilon ^b = 10$ eV.

LP1: $E_{\gamma,\mr{sh}}^{\mr{iso}} = 10 ^{51} \, \mr{ergs}$, $\Gamma_0=10$
$r={10}^{15} \, \mr{cm}$, $\xi _B =1$ and $\varepsilon^b = 100$ eV.

LP2: $E_{\gamma,\mr{sh}}^{\mr{iso}} = 10 ^{50} \, \mr{ergs}$, $\Gamma_0=10$
$r={10}^{14} \, \mr{cm}$, $\xi _B =1$ and $\varepsilon^b = 100$ eV.

Note that the model LP0 corresponds to the model FUV-ray flare (A) 
used in Murase \& Nagataki \cite{KM2}.

\section{\label{sec:leveld}Neutrino Production in GRBs}
We have assumed that the early afterglow emission comes from high
energy electrons that are accelerated via some dissipation process such as
shock dissipation. Protons also may be accelerated in internal and/or
external shocks.
If the first-order Fermi acceleration mechanism is realized,   
a proton spectrum can be written as,
\begin{equation}
\frac{d n_p}{d {\varepsilon}_{p}} = \frac{U_{p}}
{\int_{{\varepsilon}_{p}^{\mr{min}}}^{{\varepsilon}_{p}^{\mr{max}}} 
 d \varepsilon_p \left( \varepsilon_p 
\frac{dn_p}{d \varepsilon_p} \right)} \varepsilon_p^{-p},
\end{equation}
where $p$ is the spectral index. Its value is typically $p \approx 2$ in the
non-relativistic shock diffusive acceleration theory. 
In the ultra-relativistic shock limit,
$p \approx 2.2$ is obtained assuming the isotropic diffusion in the
downstream \cite{Ach1,Kes1}. In this paper, we adopt $p=2$ for the 
proton spectrum. $U_p$ is the energy density of nonthermal protons,
which is given by $U_p = \xi_{\rm{acc}}U_{\gamma}$ in the late prompt
emission model while by $U_p= \epsilon_{\rm{acc}} (E_{\rm{ej}}
/4 \pi r_{\times}^2 \Gamma_{\times}^2 \Delta_{\times})$. (Note that the
neutrino background can be calculated, given the GRB rate history. 
For this purpose, we also use the different normalization based on the 
observed UHECR flux. See the end of this subsection. ) 
The minimum energy of protons would be $\varepsilon_p^{\mr{min}} \sim$ 
a few$\times \Gamma_{\mr{rel}} m_p c^2$, although the exact value is
unknown. Here, $\Gamma_{\mr{rel}}$ is the relative Lorentz factor,
which is $\Gamma_{\mr{rel}}=\Gamma^{\prime}$ for the RS or $\Gamma_{\mr{rel}}
=\Gamma_{\mr{sh}}$ for the internal shocks. Here, we take
$\varepsilon_{p}^{\mr{min}}=10$ GeV, although the accurate value is
irrelevant for the resulting spectra. For the FS, the 
proton energy in the observer frame is given by $E_p^{\mr{min}} 
\sim$ a few $\times \Gamma m_p c^2$.

The maximum energy of cosmic-ray nuclei is determined by several criteria. 
One of the necessary conditions is obtained by comparing the Larmor 
radius of cosmic-ray nuclei with the size of the acceleration
region \cite{Rac1,Hil1}. In cases we consider, this criterion corresponds to
comparing the acceleration time scale $t_{\mr{acc}}$ with the dynamical
time scale $t_{\mr{dyn}}$. In addition, the maximum energy of
particles is limited by various cooling processes and diffusive losses
of them. Other criteria are obtained by comparing the acceleration
time scale with various cooling time scales (the synchrotron cooling
time scale $t_{\rm{syn}}$, adiabatic cooling time scale $t_{\rm{ad}}$
and so on) and with the escape time scale due to particle diffusion 
$t_{\rm{esc}}$ \cite{KM1,Rac1,Der1a}. For cosmic-ray nuclei to be
accelerated, all the above criteria should be satisfied.
As an example, let us consider the first criterion and estimate 
the possible maximum energy of protons accelerated in the FS for the 
thick ejecta colliding into ISM. We have \cite{Gal1},
\begin{eqnarray}
E_p^{\mr{max}} &\approx& Z e  \Gamma_{\times} B^{\mr{ISM}} r_{\times} 
\nonumber \\
&\approx& 5.1 \times {10}^{15} \, \mr{eV} \,
Z B_{-6}^{\mr{ISM}} E_{\mr{ej},53}^{3/8} n_0^{-3/8}  {T}_{1}^{1/8}, 
\end{eqnarray}
where $B^{\mr{ISM}}$ is the strength of the upstream magnetic field.
From the above equation, we can expect that protons cannot 
be accelerated up to ultra high energies at the FS by the first-order
Fermi acceleration mechanism \cite{Mil1,Gal1}. 
Although the second-order Fermi acceleration might allow protons to 
be accelerated up to ultra high energies \cite{Der3}, we consider 
only the first-order Fermi acceleration mechanism for the FS in this
paper. From Eq. (21), we can see that neutrinos, which are 
produced by the photomeson production process in the FS, will be negligible.
Corresponding to Eq. (21), the maximum energy of protons accelerated
in the RS is written as,
\begin{eqnarray}
E_p^{\mr{max}} &\approx& Z e B_{\times}^r r_{\times} \nonumber \\
&\approx& 2.0 \times {10}^{21} \, \mr{eV} \, Z \epsilon_{B}^{1/2} 
E_{\mr{ej},53}^{3/8} n_0^{1/8} T_1^{1/8}.
\end{eqnarray} 
Therefore, protons can be accelerated up to ultra high
energies at the RS by the first-order acceleration mechanism if other cooling
processes are not important.
On the other hand, in the late prompt emission model, the UHECR
production is typically impossible. Note that the photomeson 
cooling process can be important in very high energies in 
this model \cite{KM2}. We treat various cooling time scales properly 
in our numerical results that are shown later (see Appendix A). 

Sufficiently accelerated protons can interact with target photons 
via the photomeson production process and produce high energy pion
and muons. Neutrinos are produced via the decay of ${\pi}^{\pm} \rightarrow
{\mu}^{\pm}+{\nu}_{\mu}({\bar{\nu}}_{\mu}) \rightarrow
e^{\pm}+{\nu}_{e}({\bar{\nu}}_{e})+{\nu}_{\mu}+{\bar{\nu}}_{\mu}$. 
The neutrino production efficiency is represented by the 
photomeson production efficiency $f_{p \gamma} \equiv t_{\mr{dyn}}/t_{p
\gamma}$ (see Appendix A).
For example, let us consider the thick ejecta colliding into the ISM. In
such cases, we can approximately obtain \cite{KM2,Wax1,Wax2},
\begin{equation}
f_{p\gamma} \simeq 0.088 \frac{L_{b,48}}{r_{\times,16} {\Gamma_{\times,2}}^2
E_{10 \, \mr{eV}}^{b}} \left\{ \begin{array}{rl} 
{(E_p/E_p^b)}^{\beta-1} & \mbox{($E_p < E_{p}^{b}$)}\\
{(E_{p}/E_{p}^{b})}^{\alpha-1} & \mbox{($E _p^{b} < E_p$)} 
\end{array} \right. \label{pgamma7}
\end{equation}
where $E^b$ is the observed break energy which is either of
$\varepsilon_{\mr{ob}}^{c}$ or $\varepsilon_{\mr{ob}}^{m}$ or 
$\varepsilon_{\mr{ob}}^{m}$, and $L_b$ is the observed luminosity at
the break energy. $E_{p}^{b} \simeq 0.5 \bar{\varepsilon}
_{\Delta}m_pc^2 \Gamma_{\times}^2/E^{b}$ is the
proton break energy, where, $\bar{\varepsilon}
_{\Delta}$ is around $0.3$ GeV. For example, $E^b = 10$ eV and
$\Gamma_{\times}=100$ lead to $E_p^b \sim {10}^{20}$ eV.
In Eq. (23) we do not include the effect of multi-pion production
which is not important for the RS emission.

Next, let us consider the overlapping of the prompt emission 
with the shocked region. We can obtain (including the
effect of multi-pion production which is moderately important in very
high energies) \cite{KM2},
\begin{equation}
f_{p\gamma} \simeq 1.7 \times {10}^{-3} 
\frac{{{\Gamma}^{\prime}}^2 E_{\gamma,53}^{\mr{iso}}}
{r_{\times,16} \Delta_{0,12} 
{\Gamma}_{0,2.5}^2 E_{100 \, \mr{keV}}^{b}} \left\{ \begin{array}{rl} 
{(E_p/E_p^b)}^{\beta-1} 
\\
{(E_{p}/E_{p}^{b})}^{\alpha-1} 
\end{array} \right. 
\end{equation}
From Eq. (24), we expect that photomeson production is not so
efficient when the CBM is the ISM. If the CBM is wind-like, it becomes
more efficient \cite{Fan3}.

Finally, let us consider the late prompt emission model.
The physical conditions will be similar to those of flares. 
From Eq. (3) in Murase \& Nagataki \cite{KM2}, we have,
\begin{equation}
f_{p\gamma} \simeq 5.2 \frac{E_{\gamma,\mr{sh},50.5}^{\mr{iso}}}{r_{14.5}^2 
E_{1 \, \mr{keV}}^{b}} \left\{ \begin{array}{rl} 
{(E_p/E_p^b)}^{\beta-1} 
\\
{(E_{p}/E_{p}^{b})}^{\alpha-1} 
\end{array} \right. 
\end{equation}
From Eq. (25), we can expect that almost all the protons that are
accelerated to sufficiently high energies will be depleted due to
photomeson production. The above approximate evaluation by using
Eqs. (23-25) is good agreement with numerical results.  

We treat photomeson production in detail and calculate neutrino
spectra numerically. Generally, cooling processes of pions and
muons are important. Of course, our numerical calculations take into
account them, but the analytical consideration is convenient. 
Hence, we briefly review the neutrino emission process here. More detailed
discussions can be found in e.g., Rachen \& M\'esz\'aros \cite{Rac1}. 

One of the important cooling processes is synchrotron cooling of
pions and muons. The synchrotron break for neutrinos from pions/muons 
is determined by $t_{\pi/\mu,\mr{syn}}=\gamma_{\pi/\mu}
\tau_{\pi/\mu}$, where $\tau_{\pi/\mu}$ is the mean life times of pions and
muons. For neutrinos from pions and muons, we obtain,
\begin{subequations}
\begin{eqnarray}
E_{\nu}^{\pi,s} &\approx& \frac{1}{4} \Gamma \sqrt{\frac{6 \pi
m_{\pi}^5 c^5 }{\sigma_{\mr{T}}m_e^2 B^2 \tau_{\pi}}} \\
E_{\nu}^{\mu,s} &\approx& \frac{1}{3} \Gamma \sqrt{\frac{6 \pi
m_{\mu}^5 c^5 }{\sigma_{\mr{T}}m_e^2 B^2 \tau_{\mu}}}.
\end{eqnarray}
\end{subequations}
Above the synchrotron break, a neutrino spectrum is suppressed by
$t_{\pi/\mu, \mr{syn}}/t_{\mr{\pi/\mu}}$.  
The adiabatic break for neutrinos from pions/muons is 
determined by $t_{\pi/\mu,\mr{ad}}=\gamma_{\pi/\mu}\tau_{\pi/\mu}$.
We have,
\begin{subequations}
\begin{eqnarray}
E_{\nu}^{\pi,a} &\approx& \frac{1}{4} \Gamma \frac{t_{\pi}}{\tau_{\pi}}
m_{\pi} c^2 \\
E_{\nu}^{\mu,a} &\approx& \frac{1}{3} \Gamma \frac{t_{\mu}}{\tau_{\mu}}
m_{\mu} c^2.
\end{eqnarray}
\end{subequations}
Above the adiabatic break, a neutrino spectrum is suppressed by
$t_{\pi/\mu, \mr{ad}}/t_{\mr{\pi/\mu}}$. 

Now, we can obtain approximate neutrino spectra of 
$(\nu_{\mu}+\bar{\nu}_{\mu})$. 
For $E_{\nu}^b< E_{\nu}^{s} (<E_{\nu}^a)$, we obtain, 
\begin{equation}
E_{\nu}^2 \frac{dN_{\nu}}{dE_{\nu}} \approx \frac{1}{4} f_{p \gamma} 
E_{p}^2 \frac{dN_{p}}{dE_{p}}
\left\{ \begin{array}{ll}
{(E_{\nu}/E_{\nu}^{b})}^{\beta-1} 
\\
{(E_{\nu}/E_{\nu}^{b})}^{\alpha-1} 
\\
{(E_{\nu}^s/E_{\nu}^{b})}^{\alpha-1}{(E_{\nu}/E_{\nu}^{s})}^{\alpha-3}
\end{array} \right.
\end{equation}
For $E_{\nu}^b < E_{\nu}^{a} <E_{\nu}^s$, we have, 
\begin{eqnarray}
E_{\nu}^2 \frac{dN_{\nu}}{dE_{\nu}} &\approx& \frac{1}{4} f_{p \gamma} 
E_{p}^2 \frac{dN_{p}}{dE_{p}} \nonumber \\ 
&\times& \left\{ \begin{array}{ll}
{(E_{\nu}/E_{\nu}^{b})}^{\beta-1} 
\\
{(E_{\nu}/E_{\nu}^{b})}^{\alpha-1} 
\\
{(E_{\nu}^a/E_{\nu}^{b})}^{\alpha-1}{(E_{\nu}/E_{\nu}^{a})}^{\alpha-2} 
\\
{(E_{\nu}^a/E_{\nu}^{b})}^{\alpha-1}{(E_{\nu}^s/E_{\nu}^{a})}^{\alpha-2}
{(E_{\nu}/E_{\nu}^{s})}^{\alpha-3} 
\end{array} \right.
\end{eqnarray}
Note that the above two approximate expressions are valid as long as 
$f_{p \gamma}<1$. Although more general expressions can be obtained,
we do not show them here. We can obtain general neutrino more
quantitatively by grace of numerical calculations. 

It is important to see a lot of GRB events, because the detection of
neutrinos signals from one GRB is generally not easy \cite{Der1,KM1}.
We can calculate the neutrino background assuming the GRB rate history
(see Appendix B). In this paper, we adopt two different 
normalizations of the proton flux.
One is the normalization by the observed GRB rate and isotropic 
energy with the fixed acceleration efficiency. As noted before, we 
adopt $\epsilon_{\mr{acc}}=1/4$ for the RS model and
$\xi_{\mr{acc}}=10$ for the late prompt emission model as fiducial
values, respectively. The other normalization is based on the 
hypothesis that observed UHECRs (whose flux is $E_p^2 d \dot{N}_p/d E_p (z=0) 
\approx 0.65 \times {10}^{44} \, \rm{ergs} \, {\rm{Mpc}}^{-3} \rm{yr}^{-1}$) 
come from GRBs. When we adopt this normalization, we simply
change the normalization factor just for simplicity. For more detailed
calculations, proton spectra after the propagation in the universe 
should be fitted to the observed UHECR flux, and we have to take care of 
affections to target photon spectra by the change of 
the released energy per GRB according to the assumed local GRB rate. 
In the RS model we expect that UHECR production is possible, so that
 neutrino fluxes from the RS should
generally satisfy the Waxman \& Bahcall (WB) bound \cite{Wax3} 
because of $f_{p \gamma} \lesssim 1$ typically.
For the late prompt emission model, we adopt $\xi_{\mr{acc}}=50$ as
an optimistic value, as is noted in the previous section. 
Note that neutrino fluxes in this model do not have
to satisfy the WB bound generally (rather, it should be constrained 
by the more general bound \cite{Man1}). This is because protons 
are usually not accelerated up to ultra high energies and we can expect 
$f_{p \gamma} \gtrsim 1$ in this model. 

\section{\label{sec:level3}Numerical Results}
\subsection{\label{subsec:levele}Neutrino Spectra from GRB Early
Afterglows
}
We need to evaluate proton cooling time scales which include
the photohadronic cooling calculated by Geant4. In Fig. 1, we show the
acceleration time scale and proton cooling time scales for the model
ISM-e which represents the case of energetic ejecta that is expected
in GRB 990123. In this case, the most important cooling
process is adiabatic cooling. The maximum energy is determined by 
Eq. (22). For the ISM cases such as the models 
ISM-tc and ISM-tn, the adiabatic time scale is typically the most 
important. For the wind-like CBM cases, synchrotron cooling can
be more important. On the other hand, photohadronic cooling can
be dominant cooling process in the late prompt emission model, as 
shown in Murase \& Nagataki \cite{KM2}. Synchrotron cooling and
adiabatic cooling can also be important.

\begin{figure}[t]
\includegraphics[width=\linewidth]{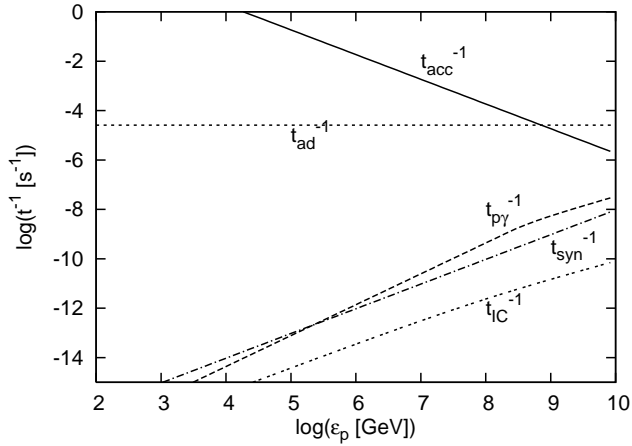}
\caption{\label{Fig1}The acceleration time scale and various cooling
time scales of protons for the model ISM-e. The energy scale is
measured in the comoving frame.}
\end{figure}     

In the case of the prompt emission, the effect of pion-multiplicity and
proton-inelasticity is moderately important in the sufficiently high energies. 
This comes from the fact that more and more target 
photons can interact with protons via double-
and multi-pion production, as the energy of incident protons becomes
high. As a result, the photohadronic cooling time scale increases with
the energy more than expected in the $\Delta$-resonance
approximation \cite{KM1}. Similar things can be applied when we 
consider the overlapping effect of the prompt emission and the late prompt emission model.  
This effect is more important for a flatter photon spectrum as demonstrated
in the previous paper. On the other hand, the RS emission 
($\alpha \sim (1.5-1.7)$ $(\mr{for} \, \varepsilon < \mr{max}[\varepsilon^{m},
\varepsilon^{c}])$ and $\beta \sim 2.2$ $(\mr{for} \, \varepsilon > 
\mr{max}[\varepsilon^{m},\varepsilon^{c}])$) has a steeper spectrum 
than that of the prompt emission ($\alpha=1$ and
$\beta=2,2$). Therefore, the multiplicity effect is less
important in the case of the RS emission and 
the $\Delta$-resonance approximation is a good
approximation. In fact, Eq. (23) is good agreement with numerical
results.

\begin{table}[t]
\begin{center}
\begin{tabular}{|c|c|c|c|c|c|c|c|}
\hline Model & $\Gamma_{0}$ & $E_{\rm{ej}}$ [ergs] &
$\Delta_{0}$ [cm] & $n$ $(A_{\ast})$ & $\epsilon_{B}^r$ & 
$\epsilon_{e}^r$ & $f_{e}^r$\\
\hline
\hline ISM-tc & ${10}^{2.5}$ & $4 \times {10}^{53}$ & $4.5 \times {10}^{11}$ & 
$5$ & 0.01 & 0.25 & 1\\
\hline ISM-tn & ${10}^{2}$ & $4 \times {10}^{52}$ & $4.5 \times 10^{11}$ & 
$0.5$ & 0.01 & 0.25& 1\\
\hline ISM-s & ${10}^{2.5}$ & $4 \times {10}^{53}$ & $4.5 \times {10}^{11}$ & 
$5$ & 0.25 & 0.25 & 0.025\\
\hline ISM-e & $300$ & $10^{55}$ & $1.05 \times {10}^{12}$ & $1$ & 
0.001 & 0.04 & 1\\
\hline ISM-eb & $300$ & $10^{54}$ & $1.05 \times {10}^{12}$ & $1$ & 0.2 & 
0.1 & 1\\
\hline \hline WIND-tc & ${10}^{2.5}$ & $4 \times 10^{53}$ & $4.5 \times 
{10}^{11}$ & $1$ & 0.01 & 0.25& 1\\
\hline WIND-tn & ${10}^{2}$ & $4 \times 10^{52}$ & $4.5 \times {10}^{11}$ 
& $0.01$ & 0.01 & 0.25& 1\\
\hline WIND-s & ${10}^{2.5}$ & $4 \times 10^{53}$ & $4.5 \times {10}^{11}$ 
& $0.1$ & 0.25 & 0.25& 0.025\\
\hline
\end{tabular}
\caption{The adopted parameter sets in the reverse-forward shock model. Note 
that the models, where the overlapping effect due to the prompt emission is 
included, are referred as ISM-tc2, ISM-s2, WIND-tc2 and WIND-s2. The 
isotropic prompt emission energy $E_{\gamma}^{\rm{iso}}={10}^{53}$ ergs is 
assumed in these four parameter sets. \label{RSFS}}
\end{center}
\end{table}
\begin{table}[t]
\begin{center}
\begin{tabular}{|c|c|c|c|c|c|}
\hline Model & $\Gamma_0$ & $E_{\gamma,\rm{sh}}^{\rm{iso}}$ [ergs] &
$r$ [cm] & $\xi_B$ & ${E^b}'$ [keV]\\
\hline
\hline LP0 & 15 & ${10}^{51.2}$ & ${10}^{15.3}$ & $1$ & 0.15\\
\hline LP1 & 10 & ${10}^{51}$ & ${10}^{15}$ & $1$ & 1\\
\hline LP2 & 10 & ${10}^{50}$ & ${10}^{14}$ & $1$ & 1\\
\hline
\end{tabular}
\caption{The adopted parameter sets in the late prompt emission model. 
\label{LP}}
\end{center}
\end{table}
\begin{figure}[t]
\includegraphics[width=\linewidth]{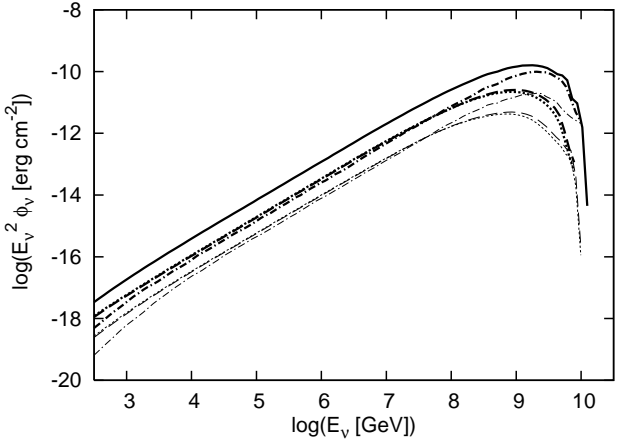}
\caption{\label{Fig2} Neutrino fluences from one energetic GRB event at
$z=1.6$. Neutrinos are produced by the decay of pion and muon whose
origins are single-pion production (thick lines), and double- and multi-pion 
production (thin lines). The model ISM-e is used. We show
fluences of muon-neutrinos $(\nu_{\mu} + \bar{\nu}_{\mu})$ 
from pion decay (dot-dashed lines), 
muon-neutrinos $(\nu_{\mu} + \bar{\nu}_{\mu})$ 
from muon decay (dashed-lines) and electron-neutrinos $(\nu_{e} +
\bar{\nu}_{e})$
from muon decay (dotted lines). The total neutrino fluence is also
shown by the thick solid line. The normalization of the proton flux is 
given by $\epsilon_{\rm{acc}}=1/4$ and neutrino oscillation is not 
taken into account.}
\end{figure}

By executing numerical calculations, we can obtain neutrino spectra 
(see Appendix A). The adopted parameter sets are summarized in Tables
I and II. First, we show the neutrino energy fluence from one GRB event.
The neutrino energy fluence can be evaluated by,
\begin{equation}
E_{\nu}^2 \phi_{\nu} = \frac{(1+z)}{4 \pi d_L^2} \Gamma
\varepsilon_{\nu}^2 \frac{d N_{\nu}}{d \varepsilon_{\nu}}, 
\end{equation}
where $\varepsilon_{\nu}$ is the neutrino energy in the comoving frame
and $d_L$ is the luminosity distance to the source for a given
cosmology. Throughout the paper, we assume the $\Lambda$CDM universe
with $\Omega_{m}=0.3$, 
$\Omega_{k}=0$, $\Omega_{\Lambda}=0.7$ and $H_0=71 \, \mr{km} 
\mr{s}^{-1} \mr{Mpc}^{-1}$.

In Fig. 2,  we show neutrino energy fluences for the model ISM-e. 
Even for prompt neutrino bursts predicted under the internal shock model, 
it is difficult to see high energy signals from one GRB event by
IceCube, which typically requires $E_{\nu}^2 \phi_{\nu} \gtrsim
{10}^{-4} \, \mr{erg} \, {\mr{cm}}^{-2}$ for 0.1 PeV neutrinos \cite{Der1}. 
Therefore, it is much more difficult to see
neutrino afterglows under the original RS model because the smaller number of 
neutrinos with higher energies is expected in this model. 

We comment on the effect of double- and multi- pion production to
neutrino spectra. In the case of the prompt emission, this effect can become 
moderately important and enhance the neutrino fluence in the very high 
energies \cite{KM1}. But this effect is hardly important in the
case of the RS emission, as seen in Fig. 2.

\begin{figure}[t]
\includegraphics[width=\linewidth]{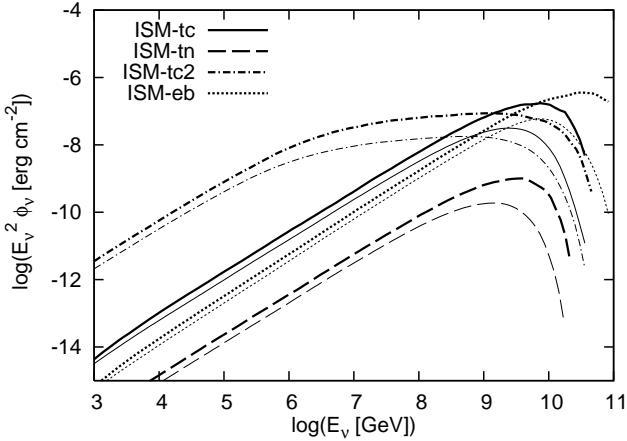}
\caption{\label{Fig3} Muon-neutrino $(\nu_{\mu} + \bar{\nu}_{\mu})$
fluences (thick lines) and electron-neutrino  $(\nu_{e} +
\bar{\nu}_{e})$ fluences (thin lines)
from one GRB event at $z=0.1$. The RS models with the ISM are
used. Note that the normalization of the proton flux is given by 
$\epsilon_{\rm{acc}}=1/4$ and neutrino oscillation is not taken into account.}
\end{figure}
 \begin{figure}[t]
\includegraphics[width=\linewidth]{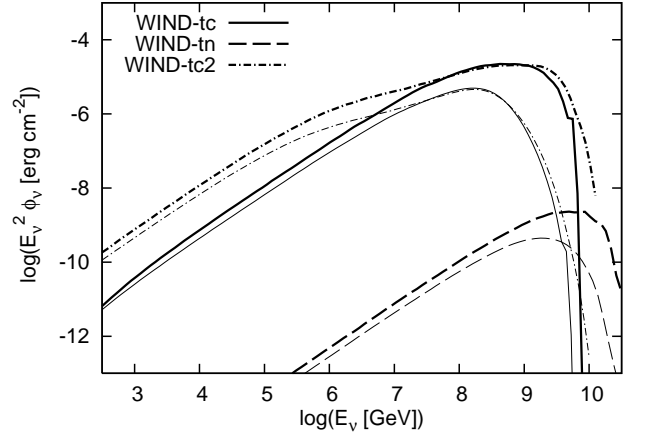}
\caption{\label{Fig4} The same as Fig. 3, but for the RS models with
the wind-like CBM.}
\end{figure}
\begin{figure}[t]
\includegraphics[width=\linewidth]{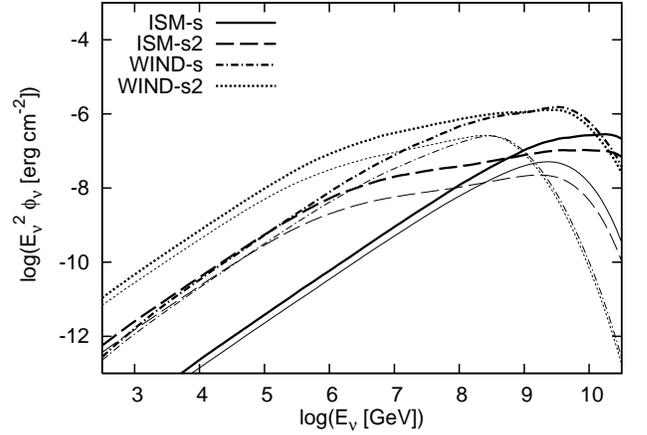}
\caption{\label{Fig5} The same as Fig. 3, but model parameters are
chosen, motivated by the modified RS model explaining the shallow
decay emission.}
\end{figure}
\begin{figure}[t]
\includegraphics[width=\linewidth]{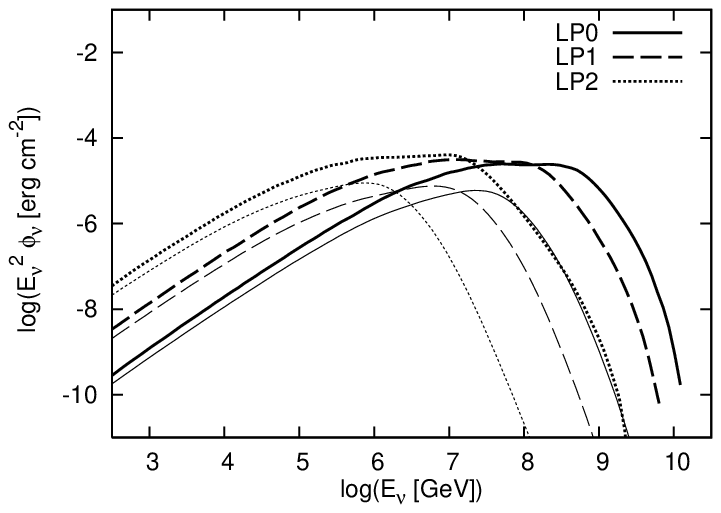}
\caption{\label{Fig6} The same as Fig. 3, but for the models based on 
the late prompt emission model. The normalization of the proton flux is given 
by $\xi_{\rm{acc}}=10$.}
\end{figure}

In Fig. 3, we show resulting spectra in the RS model with the ISM. 
For all the models, we hardly expect neutrino signals by IceCube. 
As an example, in the model ISM-tc, we have small photomeson 
production efficiency $f_{p \gamma} \sim 5 \times {10}^{-3}$ 
(at $\varepsilon_p \approx {10}^{8.5}$ GeV). In the model ISM-tn, 
smaller $E_{\rm{ej}}$ and smaller $f_{p \gamma} \sim 5 \times {10}^{-4}$ 
(at $\varepsilon_p \approx {10}^{9}$ GeV) lead to much smaller neutrino
fluences. In the original analytical prediction \cite{Wax2}, 
the neutrino energy fluence increases as neutrino energy is high. But, 
the more realistic spectra show the suppressed behavior around 
${10}^{9}$ GeV. This is just because protons have the finite 
maximum energy so that we see the cooling
effect of pions and muons in the highest energies. 
While the fluence of highest energy neutrinos is suppressed 
above $E_{\nu} \sim {10}^{9}$ GeV compared to
the analytical result, we can see 
$E_{\nu}^2 \phi_{\nu} \propto E_{\nu}^{1.2}$ for 
$E_{\nu} \lesssim {10}^{9}$ GeV, which is good agreement with 
the analytical result. In the model ISM-eb, we can see that neutrino spectra 
are extended to higher energies. This is just because the maximum energy of 
protons in this model, where large $\epsilon_B^r=0.2$ is taken, is higher than 
in the models ISM-tc and ISM-tn (by a factor of $\sim 5$ larger than in 
the model ISM-tc).    

In the model ISM-tc2, we consider overlapping of the 
prompt emission. Once this is taken into account, neutrino fluence is greatly
enhanced by two or three orders of magnitude. Therefore, the effect of this
overlapping is important. Nevertheless, the detection by IceCube will
be very difficult unless a burst with $E_{\mr{ej}}\sim {10}^{54-55}$
ergs occurs at $\sim 10$ Mpc. 

As shown in Fig. 4, we can expect that the neutrino emission occurs
efficiently in the model WIND-tc, where we obtain 
$f_{p \gamma} \gtrsim 1$ (for $\varepsilon_p \gtrsim {10}^{8}$ GeV).
It is because the crossing radius in the model WIND-tc is smaller 
than that in the models with the ISM. Below $E_{\nu} \sim {10}^{8}$ GeV, 
we can see $E_{\nu}^2 \phi_{\nu} \propto E_{\nu}^{1.2}$ (note that the
break corresponding to $\varepsilon_{\rm{ob}}^m \approx 8.3$ eV is
 $E_{\nu}^m \sim {10}^{9.8}$ GeV, which we cannot see because the
fluence is saturated at energy such that $f_{p \gamma} \sim 1$). 
While the spectrum of the model WIND-tc is in the fast cooling regime, 
that of the model WIND-tn is in the slow cooling regime.
We cannot expect high photomeson production efficiency in the latter
model. We have $f_{p \gamma} \sim 3 \times {10}^{-4}$ (at
$\varepsilon_p \approx {10}^{8.5}$ GeV) due to the smaller 
$E_\mr{ej}$ and larger crossing radius.

If we consider overlapping of the prompt emission for the models with
the wind-like CBM, neutrino fluence can be higher by one order of
magnitude due to additional target photons from the prompt emission.

In Fig. 5, neutrino spectra are calculated under the RS model with the
 higher $\epsilon_B$ and smaller $f_e$. We choose such parameters based on
 the modified RS model that explains the shallow decay emission
 \cite{Uhm1,Gen1}. We calculate for the RS produced by the head of the
 ejecta, although the RS emission would continue by the slower tail 
of the ejecta. Although additional protons supplied by the tail of the
 ejecta may produce neutrinos, our estimation on neutrino 
fluences would not be changed so much up to a factor. Target 
photon spectra in both models ISM-s and
 WIND-s, are expected in the fast cooling regime at the crossing time
 of the head of the ejecta. In this RS model, significant 
energy of the RS emission is radiated as x-ray photons that can 
interact with protons with energy $E_p \sim {10}^{9}$ GeV, 
which produce neutrinos with $E_{\nu} \sim 5 \times {10}^{7}$ GeV. 
For the model ISM-s, we have the photomeson efficiency 
$f_{p \gamma} \sim 0.01$ (at $\varepsilon_{p} \approx {10}^{8.5}$
 GeV), which is higher than that in the model ISM-tc. 
It is because photons that can interact with protons increase 
in this RS model due to its higher injection energy 
$\varepsilon_{\rm{ob}}^m \sim 0.3$ keV. 
On the other hand, in the model WIND-s,
 we obtain $f_{p \gamma} \sim 0.1$ (at $\varepsilon_p \approx {10}^{9}$
 GeV), which is smaller than that in the model WIND-tc. This is because
 most of the radiated energy is emitted as photons with 
$\varepsilon_{\rm{ob}}^m \sim 7$ keV which is higher than that in the original
 RS model. After all, it reduces the number of target 
photons that can interact with protons. Corresponding to
 $\varepsilon_{\rm{ob}}^m \sim 7$ keV, we expect the break at
 $E_{\nu}^m \sim {10}^{7}$ GeV, which can be seen in Fig. 5. 
We can also find that overlapping of the prompt emission 
enhances neutrino fluence greatly in Fig. 5.

In Fig. 6, we show results obtained under the late prompt emission
model. For all the three models, we obtain $f_{p \gamma} \gtrsim
1$. Expected muon events from one GRB event is $N_{\mu} \sim (0.3-0.6)$ events 
for the model LP2. Therefore, if bright GRBs occur, we have possibilities 
to detect neutrino signals from early afterglows in the late prompt emission 
model. Because most of the sufficiently high energy protons are depleted in 
these three models, we see the similar level of the energy fluence. In our models, 
the magnetic field strength is stronger than that in the
RS model. Hence, neutrino spectra are suppressed in the highest
energies due to cooling processes of pions and muons.
(Note that the model LP0 has the same parameter set used in Murase \&
Nagataki \cite{KM2}, but we have the higher energy fluence by a factor
 because we showed the energy fluence per internal collision 
in the previous Letter.) 

\subsection{\label{subsec:levelh}GRB Neutrino Background}
Because it is difficult to see neutrino signals from one GRB event,
 we have to observe many GRBs. Because we can find a good
 fraction of GRBs by space satellites, we expect neutrino signals 
that are coincident with bursts seen by electromagnetic
 observations. Such a correlation is important to detect 
neutrino signals from GRBs. Otherwise, neutrino signals 
may be buried below the atmospheric neutrino background and 
cosmogenic neutrino background in the high energies. 

We calculate the neutrino background from GRBs for our specific
parameter sets. 

\begin{figure}[t]
\includegraphics[width=\linewidth]{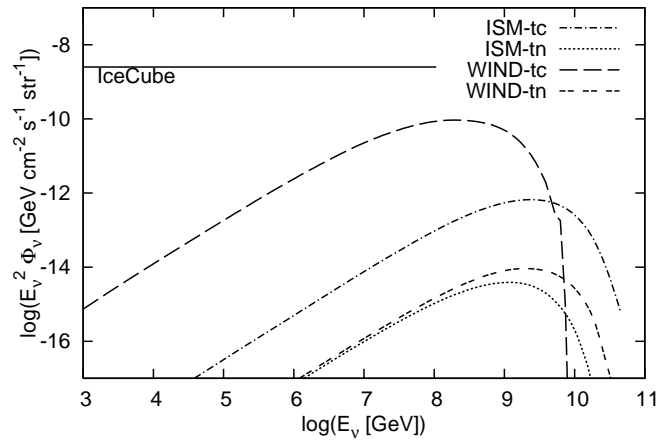}
\caption{\label{Fig7} The muon-neutrino $(\nu_{\mu}+ \bar{\nu}_{\mu})$
background from early afterglows of GRBs, based on the original
reverse-forward shock model which typically predicts infrared/optical
flashes. The normalization of the proton flux is given by
$\epsilon_{\rm{acc}}=1/4$ and neutrino oscillation is taken into account.}
\end{figure}
\begin{figure}[t]
\includegraphics[width=\linewidth]{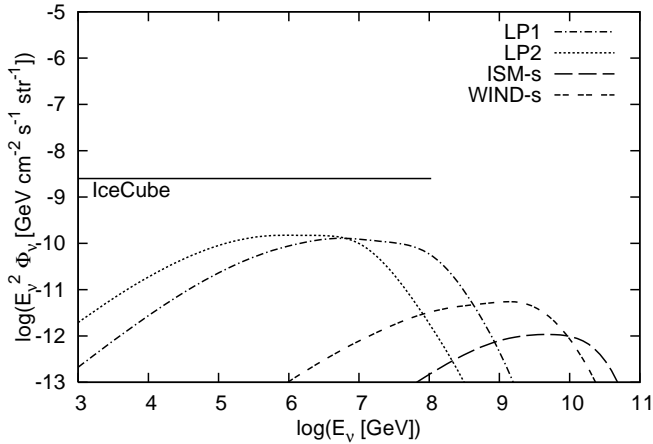}
\caption{\label{Fig8} The same as Fig. 7. But predictions are based on
the late prompt emission model and reverse-forward shock model with
the higher $\epsilon_B$ and smaller $f_e$, motivated by \textit{Swift} 
observations. The normalization of the proton flux is given by
$\xi_{\rm{acc}}=10$ for the late prompt emission model, while by 
$\epsilon_{\rm{acc}}=1/4$ for the reverse-forward shock model.}
\end{figure}
\begin{figure}[t]
\includegraphics[width=\linewidth]{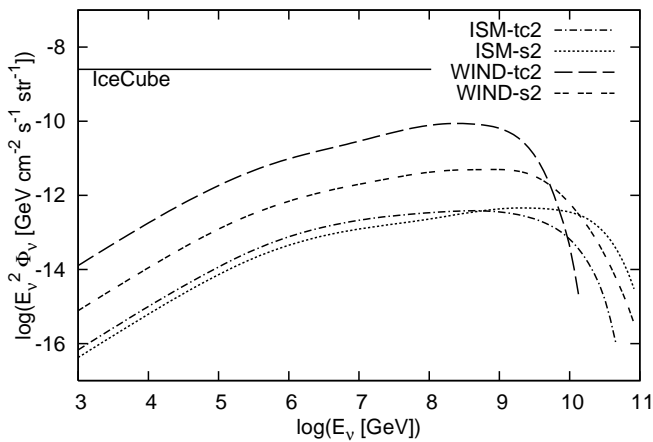}
\caption{\label{Fig9} The same as Figs. 7. But the RS model where 
the possible overlapping of the prompt emission is taken into account. The 
normalization of the proton flux is given by $\epsilon_{\rm{acc}}=1/4$.}
\end{figure}
\begin{figure}[t]
\includegraphics[width=\linewidth]{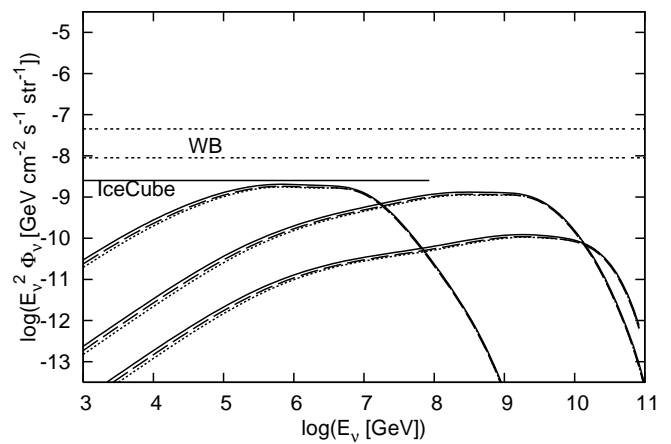}
\caption{\label{Fig10} The optimistic predictions of the neutrino 
$(\nu_{e}+ \bar{\nu}_{e} + \nu_{\mu}+ \bar{\nu}_{\mu}+ \nu_{\tau}+ 
\bar{\nu}_{\tau})$ background from early afterglows of GRBs in the 
\textit{Swift} era. The upper three curves are for the model LP2,
middle three curves for the model WIND-s2, and lower three curves for
the model ISM-s2. The dependence of the neutrino backgrounds on GRB
rate models is also shown. Solid curves are for the GRB3 model, dashed
curves for the GRB2 model, and dotted curves for the GRB1 model (see
Appendix B). In the late prompt emission model LP2, the normalization 
of the proton flux is given by $\xi_{\rm{acc}}=50$. 
In the RS models ISM-s2 and WIND-s2, the possible overlapping effect
due to the prompt emission is taken into account, and the proton flux is 
normalized to $E_p^2 d \dot{N}_p/d E_p (z=0) = 0.5 \times {10}^{44} 
\, \mr{ergs} \, \mr{Mpc}^{-3} \mr{yr}^{-1}$ which corresponds to the 
observed UHECR flux.}
\end{figure}

In Fig. 7, we show the neutrino background for the parameter sets 
based on the original RS model which was developed 
in the pre-\textit{Swift} era. Even in the model WIND-tc, expected muon 
numbers by IceCube are $N_{\mu} \sim (0.05-0.1)$ events/yr. 
It is much difficult to see signals for the other models in Fig. 7.
Even worse, if a RS is not common, neutrino signals are also even 
less expected. 

In Fig. 8, we show the neutrino background based on the late prompt
emission model and RS model where the higher $\epsilon_B$ and 
smaller $f_e$ are assumed.
Both are developed for explanation of the shallow decay emission.
In the late prompt emission model, we can expect that the optical
depth for photomeson production is high enough. In such cases, although
UHECR production is difficult, GRBs are efficient neutrino emitters. 
We obtain $N_{\mu} \sim (0.5-1)$ events/year for
the model LP1 and $N_{\mu} \sim (1-3)$ events/year for
the model LP2, respectively. Although such signals can marginally be
detected by IceCube, the significant detection requires  
neutrino detectors larger than IceCube.  
For the RS model, we have $N_{\mu} \sim (0.002-0.005)$
events/year for the model WIND-s.

In Fig. 9, we show the neutrino background in the RS models 
where the possible overlapping of the prompt emission is taken into account. The
effect of overlapping enhances neutrino fluxes greatly below $\sim
100$ PeV.

In Fig. 10, we adopt the different normalization of the proton flux
from Figs. 7-9. For the RS models, it comes from the hypothesis that
the observed UHECRs come from the RS. Note that large baryon
load is assumed for all the curves in Fig. 10. In addition, we have
also assumed that all the GRBs are in thick ejecta regime for the RS
models. Although such assumptions may be optimistic, they can lead to
detectable neutrino signals from early afterglows by future observations.  
The dependence of the neutrino backgrounds on GRB rate models is also
shown. We can see that the resulting neutrino backgrounds are not
senstive to adopted GRB rate models GRB1-GRB4 (where curves for the
model GRB4 are not shown, but they are also similar to the other
curves). 

\begin{figure}[t]
\includegraphics[width=\linewidth]{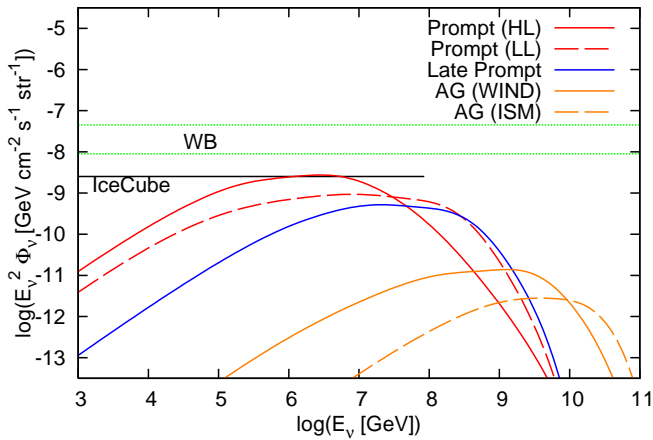}
\caption{\label{Fig11} The fiducial predictions of the 
neutrino $(\nu_{e}+ \bar{\nu}_{e} + 
\nu_{\mu}+ \bar{\nu}_{\mu}+ \nu_{\tau}+ \bar{\nu}_{\tau})$ background
from GRBs in the \textit{Swift} era. Prompt (HL): prompt neutrino bursts
from cosmological (high luminosity) GRBs; $E_{\mr{GRB}}=1.24 \times
{10}^{51}$ ergs, $E_{\gamma,\mr{sh}}^{\mr{iso}} = {10}^{51} \, \mr{ergs}$, 
$r= {10}^{13-14.5} \, \mr{cm}$ and $\Gamma_0={10}^{2.5}$, $\xi _{B}=1$
and $\xi _{\mr{acc}}=10$ \cite{KM1,Ach2}.
Prompt (LL): prompt neutrino bursts from low luminosity GRBs; the
local observed rate $\rho_{\mr{LL}}(z=0)=500 \, \mr{Gpc}^{-3}
\mr{yr}^{-1}$, $E_{\gamma}^{\mr{iso}} \simeq {10}^{50} \,
\mr{ergs}$, $r= {10}^{15} \, \mr{cm}$, $\Gamma_0=10$, $\xi _{B}=1$
and $\xi _{\mr{acc}}=10$ \cite{KM3}. 
Late Prompt: flaring neutrino flashes and/or neutrino early afterglows
under the late prompt emission model; 
$E_{\mr{LP}}=0.1 E_{\mr{GRB}}$, $L_{b}={10}^{48} \, \mr{ergs} \,
\mr{s}^{-1}$,  $r= {10}^{15.3} \, \mr{cm}$, $\Gamma_0=15$, $\xi _{B}=1$
and $\xi _{\mr{acc}}=10$ (the model LP0) \cite{KM2}. AG (WIND): neutrino early
afterglows under the reverse-forward shock model with the wind-like CBM; 
the model WIND-s is
assumed (see the text). AG (ISM): neutrino early afterglows under
the reverse-forward shock model with the ISM; the model ISM-s is
assumed (see text). WB: Waxman-Bahcall bounds \cite{Wax3}. 
$\xi _{B}$ and 
$\xi_{\mr{acc}}$ are defined as $\xi _{B} \equiv U_{B}/U_{\gamma}$ and
$\xi _{\mr{acc}} \equiv U_{p}/U_{\gamma}$, respectively. 
For the fast cooling case and the proton acceleration efficiency
$\zeta_p \sim 1$, we have $\xi _{B} \sim (\epsilon _{B}/\epsilon
_{e})$ and $\xi_{\mr{acc}} \sim 1/\epsilon _{e}$, where $\epsilon_e$ is the
fraction of the internal energy density carried by electrons.
In all the scenarios, the GRB3 model is used (see Appendix B).
}
\end{figure}
\begin{figure}[t]
\includegraphics[width=\linewidth]{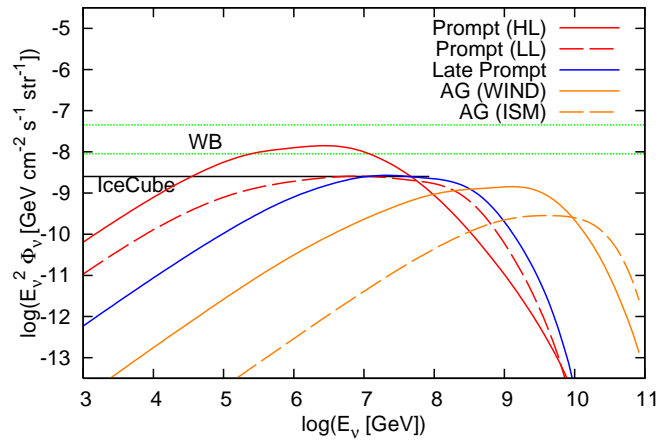}
\caption{\label{Fig12} The same as Fig. 11, but the optimistic
predictions are shown. $\xi_{\mr{acc}}=50$
is used for Prompt (HL) and Late Prompt. For Prompt (LL), 
AG (WIND) and AG (ISM), the proton flux is normalized to 
$E_p^2 d \dot{N}_p/d E_p (z=0) = 0.5 \times {10}^{44} 
\, \mr{ergs} \, \mr{Mpc}^{-3} \mr{yr}^{-1}$ which corresponds to the
 observed UHECR flux.}
\end{figure}

In Figs. 11 and 12, we show various predictions on 
the neutrino background from GRBs, motivated by observations 
in the \textit{Swift} era. 
Fig. 11 represents our fiducial predictions. For cosmological (high
luminosity (HL)) GRBs, we can expect $N_{\mu} \sim (6-11)$ events/yr,
which is consistent with previous works \cite{Wax1,Gue1}. For low
luminosity (LL) GRBs \cite{KM3,Gup1}, 
we can expect $N_{\mu} \sim (2-5)$ events/yr,
although neutrino signals coincident with LL GRBs are usually not expected by 
\textit{Swift} and GLAST. For flares \cite{KM2} and early afterglows under the
late prompt emission model (the model LP0), we can expect 
$N_{\mu} \sim (0.4-1)$
events/yr. In the very high energy region of the neutrino background, 
neutrinos from LL GRBs, flares and
early afterglows will be dominant to those from HL GRBs. 
For AG (WIND) (the model
WIND-s) and AG (ISM) (the model ISM-s), we expect $N_{\mu} 
\sim (0.002-0.005)$ events/yr and $N_{\mu} \sim 
(0.0002-0.0004)$ events/yr, respectively. 
    
Fig. 12 expresses the optimistic predictions.
For HL GRBs, we show the neutrino background for $\xi_{\rm{acc}}=50$. 
The optimistic model for $\xi_{\rm{acc}}=100$ is being constrained by present and 
future-coming neutrino observations \cite{Ach2,Ach3}. Note that the 
neutrino flux is allowed to exceed the WB bound in principle if the 
collision radii are small enough, but the neutrino
flux for Prompt (HL) would typically not exceed the WB bound
\cite{KM1,Wax3,Man1}. For Prompt (LL), we normalize the proton flux to 
$E_p^2 d \dot{N}_p/d E_p (z=0) = 0.5 \times {10}^{44} \, \mr{ergs} \, 
\mr{Mpc}^{-3} \mr{yr}^{-1}$ just for comparison. Although the shown 
parameter set allows protons to be accelerated up to the ultra high 
energy $E_p \sim 4 \times {10}^{20}$ eV, the explanation of the 
observed UHECRs might be difficult if UHECRs are all protons
\cite{KM3}. (But note that heavier nuclei could be accelerated up to
the highest energies.) If accelerated protons cannot achieve the ultra high
energies (below $\sim {10}^{19}$ eV), the neutrino flux does not have
to satisfy the WB bound in principle 
and could exceed the flux shown in Fig. 12 if the larger baryon load or
higher rate is possible.  
For Late Prompt, UHECR production is typically impossible, and we just adopt 
$\xi_{\mr{acc}}=50$. For the RS models, 
we normalize the proton flux to the observed UHECR flux. We have 
$N_{\mu} \sim (0.2-0.5)$ events/yr for AG (WIND) and $N_{\mu} \sim
(0.02-0.05)$ events/yr for AG (ISM).   

\section{\label{sec:level4}Summary and Discussion}
In this paper, we have studied the neutrino emission from early afterglows
of GRBs under the assumption that baryon acceleration occurs at the
shocks such as the RS and late internal shocks.
With acceptable parameters, we have shown that neutrino signals 
from early afterglows could marginally
be detected by IceCube under the late prompt emission model,
while they are not expected under the RS model. Hence, the neutrino
detection from early afterglows will be likely to suggest that neutrinos
come from the late prompt emission (including flares). Future neutrino 
telescopes larger than IceCube are beneficial, although the
simultaneous electromagnetic multi-wavelength observations are
indispensable. 

Our conclusions are summarized below.

(1) We have revisited the neutrino emission from the RS. We have also
taken into account
the cross section of photomeson production quantitatively 
without using the $\Delta$-resonance approximation.  
The effect of pion-multiplicity and proton-inelasticity 
is not important in th RS model and the $\Delta$-resonance 
approximation is a good approximation.
The neutrino flux is suppressed in the highest energies 
because of the finite proton's maximum
energy and cooling of secondary particles.
In the original RS model, it is very difficult to detect neutrino
signals for our fiducial parameter sets, even when all the GRBs accompany the
RS. In addition, recent observations imply the lack of
infrared/optical flashes. If it is attributed to the intrinsically
weak RS, we can expect neutrino production in the RS region only for a
fraction of GRBs with the sufficiently strong RS.  

(2) One of the suppression mechanisms for infrared/optical flashes 
is the overlapping of the prompt emission. In this scenario, the RS
occurs, but cooling of electrons suppresses the emission in the
infrared/optical bands. We have also taken into account this effect,
which can dramatically enhance neutrino
fluxes. In the case of the thick ejecta colliding into the ISM,
enhancement of the flux by about two or three orders of magnitude 
is expected, while by one order of magnitude in the case of the thick
 ejecta colliding into the wind-like CBM. Expected muon events 
can increase by a factor. Although this component may
be hidden by prompt neutrino bursts, such contribution will be
important in the sense that we do not have to assume the proton 
acceleration in the region where the prompt emission occurs.

(3) Recently, it is suggested that the shallow decay emission in early
afterglows may be explained by the modified RS model. As such an example, 
we have estimated the neutrino flux under the RS model with $f_e < 1$.  
If this picture is true, we expect that the RS is common for all the
GRBs. Therefore, we can cumulate the neutrino flux 
from each burst and obtain the neutrino background. We expect muon
events by IceCube, $N_{\mu} \sim (0.002-0.005)$ events/yr in the model
WIND-s with the moderate baryon load. The hypothesis that the observed
UHECRs come from the RS leads to the higher neutrino flux, but it
requires the large baryon load. 

(4) One of the recently suggested models is the late prompt emission 
model. The physical conditions are similar to those in flares. With acceptable
parameters, neutrino production can be efficient because the emission 
occurs at smaller radii in this model. We can have 
$f_{p \gamma} \gtrsim 1$, which could allow us to estimate the
proton flux from the neutrino flux. In this model, we could expect
detectable muon events by IceCube, although the larger telescopes
would be desirable for sufficiently significant detections. If we see 
such signals, neutrinos will provide us with useful information 
on baryon acceleration in GRBs and one of the clues to the model of
early afterglows. 

(5) From Figs. 11 and 12, we have seen that the contribution from 
Prompt (HL) to the neutrino background may be the most important 
below $\sim 10$ PeV, while the neutrino background from 
early afterglows can be more important in the very high energies
$\gtrsim 10$ PeV. Neutrino signals from early afterglows are expected to 
be coincident with the early afterglow emission and give us additional chances 
to detect high energy neutrinos. However, note that the photomeson 
production efficiency $f_{p \gamma}$ is sensitive to collision radii 
$r$ \cite{KM1}. For Prompt (HL) with $\Gamma_0 \gtrsim 100$, we have typically 
${\rm{min}}[f_{p\gamma},1] \sim (0.1-1)$, and the neutrino background can be smaller 
than that shown in Figs. 11 or 12 where we have $f_{p\gamma} \sim 0.5$ 
effectively (see, e.g., the curves for the 
parameter set B used in Murase \& Nagataki \cite{KM1}). 
On the other hand, we obtain typically $f_{p\gamma} \gtrsim 1$ 
for small $\Gamma_0 \sim 10$, which is expected in case of the 
late prompt emission model with $E_{\rm{LP}} \sim 0.1 E_{\rm{GRB}}$.
Therefore, the neutrino background from early afterglows in the late prompt 
emission model can be comparable to that from the prompt emission. Roughly 
speaking, the three possibilities (Prompt (HL), Prompt (LL) and Late Prompt) 
can give comparable contributions to the diffuse neutrino background.

It is important to see neutrinos associated with gamma-rays. 
For HL GRBs, flares and early afterglows, we can expect conincidence, while
not for LL GRBs except for very nearby events. 
For the short-lived RS models where the overlapping of the prompt emission occurs, 
it will be difficult to distinguish between RS neutrinos and 
prompt neutrinos. This is because the short-lived
RS emission for the thick ejecta case lasts for the duration of bursts 
$\sim T$. 
If the long-lived RS emission occurs, the long-term neutrino emission will be
also expected. However, it may be contaminated by neutrino flashes 
from flares or the long-term neutrino emission from the FS. The latter could
occur if protons are accelerated up to very high energies 
by other mechanisms such as the second-order Fermi acceleration 
\cite{Der2,Der1a,Li1}.   

Our predictions are for specific parameter sets.
Of course, parameters such as $E^b$, $E_{\gamma}^{\mr{iso}}$ have
dispersion. A possible wide range variability may also suggest dispersion of 
$E_{\gamma,\mr{sh}}^{\mr{iso}}$. Some of the parameters might be related
like the $E^b$-$E_{\gamma}^{\mr{iso}}$ relation for the prompt emission.
More comprehensive studies are needed in future, and more and more 
observations would enable us to take into account 
the distribution of parameters.

So far, the amount of protons has not been well constrained from
observations. In Fig. 11, we have shown the relatively moderate cases, 
while the optimistic cases are shown in Fig. 12. Due to the lack of 
knowledge on the collisionless shocks and particle acceleration, we cannot say 
whether such optimistic choices are possible or not.  
But note that too large values might be implausible from recent
observations \cite{Zha5}. Unless there is significant missing energy
(which might exist, for example, if accelerated protons carry away
significant energy), high radiative efficiency of GRBs 
implies $\epsilon_e \sim (0.1-1)$, which leads to $\xi_{\mr{acc}}
\lesssim 10$. On the other hand, the UHECR hypothesis and recently
estimated local GRB rate require the large baryon load for both of the
prompt and RS scenarios. 
We have assumed that nonthermal protons are contained,
which is expected from the internal or external shock model. However, 
this might not be true if the outflow is Poynting-dominated. 
The prompt or RS emission may come from magnetic dissipation processes 
such as reconnection, where much baryons do not have to be contained. 
Future neutrino observations are fruitful, because they could 
provide us with not only information on the physics of GRB outflows but also 
one of the clues to unknown baryon load and acceleration in GRBs. 

In this paper, we have focused on high energy neutrinos from
GRBs. Generally, the high gamma-ray emission from proton-induced
components such as muon synchrotron and secondary pair synchrotron 
should accompany neutrinos. Such high energy gamma-rays can appear 
after complicated pair-photon cascades. Theoretically, 
efficient proton acceleration might induce distinctive GeV-TeV
components and such possibilities are studied by several authors
\cite{Bot1,Der4,Der5,Pee1,Zha6}. 
If the overlapping of the prompt emission is common, we would expect 
GeV-TeV flashes from the RS via the electron inverse-Compton scattering
process \cite{Bel1}. In addition, the high energy emission via this 
process is predicted under the
several early afterglow models motivated by
\textit{Swift} observations \cite{Fan5,Wan1}. 
Future-coming GLAST observations \cite{Mce1} 
would also be useful for testing the models.

\begin{acknowledgments}
K.M. thanks the referees, S. Nagataki, S. Yoshida, K. Ioka, K. Asano 
and K. Toma for very profitable comments. 
The work of K.M. is supported by a Grant-in-Aid for JSPS Fellows.
\end{acknowledgments}

\appendix
\section{\label{sec:level6}Method of Calculation}
We briefly describe our method of calculation. The method is
essentially the same as that used in Murase \& Nagataki \cite{KM1,KM2}.

The most promising acceleration process of radiating particles is the Fermi 
acceleration mechanism. Electrons and protons are usually 
accelerated in GRBs by the Fermi acceleration mechanism. 
The first-order Fermi acceleration occurs by diffusive scattering 
of particles across strong shocks. For non-relativistic shock
acceleration, the acceleration time scale is,  
\begin{eqnarray}
t_{\mr{acc}} \approx \frac{3r_{\mr{c}}}{r_{\mr{c}}-1}
\frac{1}{U_1^2}(\kappa_{\mr{u}}+r_{\mr{c}} \kappa_{\mr{d}}),
\end{eqnarray}
where $r_\mr{c}$ is the compression ratio, $U_1$ is the velocity of
the upstream fluid and $\kappa_{u,d}$ is the diffusion coefficients. 
In the ultra-relativistic shock limit, with the assumption on the 
sufficient amplification of the downstream magnetic field, we can
obtain the acceleration time as \cite{Gal1}, 
\begin{equation}
t_{\mr{acc}} \sim \frac{\varepsilon_p}{\sqrt{2} eB \Gamma_{\mr{rel}}c} \,
\mr{max}\left [1,{\left(
\frac{\sqrt{2} \Gamma_{\mr{rel}}l_{\mr{coh}}}{r_{L}} \right
)}^{-1} \right ],
\end{equation}
where $\Gamma_{\mr{rel}}$ is the relative Lorentz factor between
fluids, $l_{\mr{coh}}$ is the coherent length of the upstream 
magnetic field and $r_L$ is the Larmor radius.   
Let us assume the Bohm diffusion coefficient and sufficiently
large coherent length of the upstream magnetic field. Then, we can 
write the acceleration time of protons as follows,
\begin{equation}
t_{\mr{acc}}\equiv \eta \frac{r_{L}}{c}=\eta\frac{\varepsilon
_{p}}{eBc} \label{acc}
\end{equation}
For the case of mildly relativistic ($\Gamma_{\mr{rel}} \sim$  a few) 
shocks, $\eta \sim 1$ may be possible. 
The acceleration time scale for the second-order Fermi 
acceleration is given by,
\begin{equation}
t_{\mr{acc}} \sim \left( \frac{r_{L}}{c {\beta_A}^{2}} \right)
{\left( \frac{B}{\delta {B}} \right)}^{2},
\end{equation}
where ${\beta}_{A} \equiv v_{A}/c$, $v_{A}$ is Alfv\'en velocity 
and $\delta B$ is the strength of the turbulence in the 
magnetic field \cite{Kul2,Rac1}. The limit $\beta_A =1$ and
$\delta B=B$ which could be expected in the downstream of relativistic
shocks also leads to the similar expression to Eq. (A3).  

Proton's maximal energy is also constrained by various cooling 
processes. In this paper, we treat proton synchrotron cooling, 
inverse Compton (IC) cooling, adiabatic cooling and photohadronic
cooling.
First, the synchrotron loss time scale for relativistic protons is,
\begin{equation}
t_{\mr{syn}}=\frac{3m_{p}^4 c^3}{4 \sigma _{T}
m_{e}^2}\frac{1}{\varepsilon_{p}}\frac{1}{U_{B}} \label{sync}.
\end{equation} 
Second, the IC cooling time scale is given by \cite{Jon1},
\begin{equation}
t_{\mr{IC}}^{-1}=\frac{c}{2 \gamma _{p}^{2}} \left(
\frac{m_{e}^{2}}{m_{p}^{2}} \right) \pi r_{0}^{2} m_{p}^{2} c^{4} \int
_{0}^{\infty} d \varepsilon \varepsilon ^{-2} \frac{dn}{d \varepsilon}
\frac{F(\varepsilon,\gamma_p)}{\beta _{p} (\gamma _{p}-1)} 
\end{equation}
where,
\begin{eqnarray}
F(\varepsilon,\gamma) &\equiv& \gamma [f_1 (z_a) - f_1 (z_b)]  \nonumber \\
&-& (\varepsilon/m_p c^2) [f_2 (z_a) - f_2 (z_b)] \nonumber \\
z_a &\equiv& \frac{\varepsilon}{m_p c^2}(\gamma+ \sqrt{\gamma^2 -1})
\nonumber \\
z_b &\equiv& \frac{\varepsilon}{m_p c^2(\gamma+ \sqrt{\gamma^2 -1})}
\nonumber \\
f_1(z) &\equiv& (z+6+3/z) \ln (1+2z) \nonumber \\
&-& (22z^3/3 + 24 z^2 + 18z +4) \nonumber \\
&\times& {(1+2z)}^{-2} - 2 + 2 \mr{Li}_2(-2z) \nonumber \\
f_2 (z) &\equiv& (z+31/6+5/z+3/2z^2) \ln (1+2z) \nonumber \\ 
&-& (22z^3 /3 + 28 z^2 + 103z +17 + 3/z) \nonumber \\
&\times& {(1+2z)}^{-2} - 2 + 2 \mr{Li}_2(-2z) \nonumber \\
\mr{Li}_2(z) &=& - \int_0^z dz^{\prime} \frac{\ln
(1-z^{\prime})}{z^{\prime}} \,\,\, (\mbox{for complex }z) \nonumber \\
\mr{Li}_2(z) &=& \sum_{n=1}^{\infty} \frac{z^n}{n^2} \,\,\,
(\mbox{for} \, \, |z|<1). \nonumber
\end{eqnarray}
Third, the photohadronic cooling time scale is,
\begin{equation}
t^{-1}_{p\gamma} = \frac{c}{2{\gamma}^{2}_{p}} 
\int_{\bar{\varepsilon}_{\mr{th}}}^{\infty} \! \! \! d\bar{\varepsilon} \, 
{\sigma}_{p\gamma}(\bar{\varepsilon}) {\kappa}_{p}(\bar{\varepsilon})
\bar{\varepsilon} \int_{\bar{\varepsilon}/2{\gamma}_{p}}^{\infty} \! \! \! \! \! \! \! \! \! d \varepsilon \, {\varepsilon}^{-2} \label{pcool}
\frac{dn}{d\varepsilon}.
\end{equation}
where $\bar{\varepsilon}$ is the photon energy in the rest frame of
proton, $\gamma _{p}$ is the proton's Lorentz factor, $\kappa _{p}$ is
the proton-inelasticity, and $\bar{\varepsilon} _{\mr{th}}$ is the
threshold photon energy for the photohadronic process in the rest frame of
the incident proton. In sufficiently high energies, photomeson cooling
process is much important, where the threshold energy is 
$\bar{\varepsilon}_{\mr{th}} \approx 145 \, \mr{MeV}$. We calculate
Eq. (A7) by using Geant4 whose total cross section is fairly good agreement
with experimental data \cite{Ago1}, and the calculated mean free path
is also good agreement with that obtained by the other code \cite{Tak1,Muc1}.  
Fourth, we take into account the adiabatic cooling process, which has
a time scale $t_{\mr{ad}}$ independent of the proton energy. In fact, 
direct ejection of protons from the emission region may 
depend on a proton energy if diffusive losses are relevant. For 
simplicity, we neglect such diffusive losses and assume that protons are
confined over the time scale set by adiabatic expansion. When
the fluid is relativistic, we have,  
\begin{equation}
t_{\mr{ad}}^{-1}=\frac{1}{3}(\bnabla \cdot \mathvec{V}) \sim t_{\mr{dyn}}^{-1},
\end{equation}
where $\mathvec{V}$ is the fluid velocity.
From above time scales, the total proton loss time 
scale is expressed as $t_{p}^{-1} \equiv t_{p\gamma}^{-1} + t_{\mr{syn}}^{-1} +
t_{\mr{IC}}^{-1} + t_{\mr{ad}}^{-1}$. The proton's maximum energy can be
determined by $t_{\mr{acc}} < t_p$.

Accelerated protons interact with target photons via photomeson
production. Pion spectra can be obtained by executing Geant4, which
are written as, 
\begin{eqnarray}
\frac{dn_{\pi}}{d\varepsilon _{\pi} dt} &=& \int _{\varepsilon _{p}^{\mr{min}}}^{\varepsilon _{p}^{\mr{max}}} \!\!\!\!\!  
d \varepsilon _{p} \, \frac{d n_{p}} {d \varepsilon _{p}} \int _{\varepsilon ^{\mr{min}}}^{\varepsilon ^{\mr{max}}} \!\!\!\!\! 
d \varepsilon \, \frac{dn}{d\varepsilon} \int \frac{d \Omega}{4\pi} \nonumber \\
&\times& \frac{d \sigma _{p\gamma}(\varepsilon,\Omega,\varepsilon
_{p}) \xi _{\pi}}{d\varepsilon _{\pi}} \, c, \label{getpion}
\end{eqnarray}
where $d n_{p}/d \varepsilon _{p}$ and $d n/d \varepsilon$ is proton
and photon distribution in the comoving frame, $\xi _{\pi}$ is the
pion-multiplicity. Owing to Geant4, we can include
the effects of proton-inelasticity and pion-multiplicity. Although 
the treatments of photomeson production are greatly
improved by using Geant4, compared to the $\Delta$-resonance 
approximation which is often used by other authors, 
Geant4 has some problems in parametrization \cite{KM1}. Therefore, we adopt 
the more improved approximate treatment \cite{KM2,KM3}. 
Below 3 GeV, we use the experimental data \cite{PDG1,Sch1}
(in the previous paper \cite{KM1}, we approximated
${\pi}^{+}:{\pi}^{0}=1:1$ for single-pion production, and 
${\pi}^{+}{\pi}^{-}:{\pi}^{+}{\pi}^{0}=7:4$ for double-pion production 
\cite{Sch1,Rac1}. Both approximations do not change our
conclusions). Above 3 GeV, we use the Geant4 approximation. 
Such an approximation is sufficient for astrophysical applications 
to calculation of GRB neutrinos, and our results for neutrino
spectra are quantitatively improved compared to most of the previous
works, where the $\Delta$-resonance approximation is used.

We can obtain neutrino spectra from the well-known decay kinematics. 
Neutrinos are produced by the decay 
of ${\pi}^{\pm} \rightarrow {\mu}^{\pm}+{\nu}_{\mu}({\bar{\nu}}_{\mu}) 
\rightarrow e^{\pm}+{\nu}_{e}({\bar{\nu}}_{e})+{\nu}_{\mu}+{\bar{\nu}}_{\mu}$. 
The life times of pions and muons are $t_{\pi}={\gamma}_{\pi}
{\tau}_{\pi}$ and $t_{\mu}={\gamma}_{\mu} {\tau}_{\mu}$,
respectively. Here, ${\tau}_{\pi}
=2.6033 \times 10^{-8} \mr{s}$ and ${\tau}_{\mu}=2.1970 \times 10^{-6} 
\mr{s}$ are the mean life times of each particle. 
When pions decay with the spectrum $dn_{\pi}/d\varepsilon _{\pi}$, 
by ${\pi}^{\pm} \rightarrow
{\mu}^{\pm}+{\nu}_{\mu}({\bar{\nu}}_{\mu})$, the spectrum 
of neutrinos is given by \cite{Sch2},
\begin{equation}
\frac{dn_{\nu}}{d\varepsilon _{\nu}}=\frac{m_{\pi}c}{2\varepsilon _{\nu}^{*}} \int ^{\infty}_{\varepsilon _{\pi}^{\mr{min}}} \! \! \! \! \!
d\varepsilon _{\pi} \, \frac{1}{p_{\pi}} \frac{dn_{\pi}}{d\varepsilon _{\pi}}
\end{equation}
where, $\varepsilon_{\nu}^{*}=\frac{(m_{\pi}^2-m_{\mu}^2)c^2}{2m_{\pi}}$,
$\varepsilon _{\pi}^{\mr{min}}= \frac{(\varepsilon_{\nu}^{*}
/\varepsilon _{\nu}+\varepsilon _{\nu}/\varepsilon_{\nu}^{*})m_{\pi}c^2}{2}$.
Similarly, we can get the muon spectrum from the pion spectrum. Muon decay 
is the three-body-decay process, which is slightly more complicated 
than the case of two-body-decay. Given the spectrum of muon, it can be 
calculated by the following equation \cite{Sch2}, 
\begin{eqnarray}
\frac{dn_{\nu}}{d\varepsilon _{\nu}}=\int
_{\varepsilon_{\mu}^{\mr{min}}}^{\infty} 
\! \! \! \! \! d\varepsilon _{\mu} \!\!\! \! \! \! \! && \frac{1}{cp_{\mu}} 
\frac{d n_{\mu}}{d\varepsilon _{\mu}} 
\int _{\varepsilon _{\nu 1}^{*}}^{\varepsilon _{\nu 2}^{*}} \! \! \! 
d \varepsilon _{\nu}^{*} \, 
\frac{1}{\varepsilon _{\nu}^{*}} \nonumber \\ 
&& \times (f_0(\varepsilon _{\nu}^{*}) \mp \mr{cos} 
\theta _{\nu}^{*} f_1(\varepsilon _{\nu}^{*}))
\end{eqnarray}
where, $\varepsilon _{\nu 1}^{*}=\gamma _{\mu}\varepsilon _{\nu} - 
{(\gamma _{\mu}^{2}-1)}^{1/2}\varepsilon _{\nu}$,
$\varepsilon _{\nu 2}^{*} = \mr{min}
[\gamma _{\mu}\varepsilon _{\nu} + 
{(\gamma _{\mu}^{2}-1)}^{1/2}\varepsilon _{\nu}, \, 
(m_{\mu}^2-m_{e}^2)c^2/2m_{\mu}]$,
${\varepsilon_{\nu}^{*}}=\frac{m_{\mu}^2 -m_{e}^2}{2m_{\mu}}c^2$
that are defined in the muon rest frame and
and for muon neutrinos, $f_0(x)=2x^2(3-2x)$, $f_1 (x)=2 x^2 (1-2x)$
and for electron neutrinos, $f_0(x)=12x^2(1-x)$, $f_1 (x)=12 x^2 (1-x)$, 
where $x \equiv 2\varepsilon _{\nu}^{*}/m_{\mu}c^2$ and
$\theta_{\nu}^{*}$ is the angle between the muon spin and the
direction of a neutrino. 

Because of cooling processes of ${\pi}^{\pm}$ and
${\mu}^{\pm}$, we have to apply Eqs. (A10) and (A11) at each time
step, when we solve the following equation,
\begin{eqnarray}
\frac{\pd}{\pd t} \left( \frac{dn_{\pi,\mu}}{d
{\varepsilon}_{\pi,\mu}}({\varepsilon}_{\pi,\mu},t) \right) + 
\frac{\pd}{\pd {\varepsilon}_{\pi,\mu}} \left(
\dot{{\varepsilon}}_{\pi,\mu}
\frac{dn_{\pi,\mu}}{d{\varepsilon}_{\pi,\mu}}
({\varepsilon}_{\pi,\mu},t) \right) \nonumber \\ = -\frac{1}{t_{\pi,\mu}} 
\frac{dn_{\pi,\mu}}{d{\varepsilon}_{\pi,\mu}}({\varepsilon}_{\pi,\mu},t)
+ Q_{\pi,\mu} ({\varepsilon}_{\pi,\mu},t), \,\,\,\,\,\,\,\,\,\,
\end{eqnarray}
where $Q_{\pi,\mu}$ represents the source term of pions and
muons due to photomeson production and decay of pions, respectively. 
The synchrotron cooling time scale is given by 
replacing proton mass with pion or muon mass in Eq. (A6). 
The adiabatic cooling time scale is still comparable to 
dynamical time scale. We also treat the IC process 
including the Klein-Nishina effect for pions and muons. We neglect the
$\pi \gamma$ process (including the photomeson production and photopair
production), because other cooling processes are usually 
more important in our interested cases. 
Throughout the calculations, we also neglect neutrinos from 
neutron decay 
$n \rightarrow p+e^{-}+{\bar{\nu}}_{e}$, whose time scale is usually 
much larger than the dynamical time scale $t_{\mr{dyn}}$ for the 
short-lived emission (but note that these
components will also contribute to the diffuse neutrino background). 

\section{\label{sec:level7}GRB Neutrino Background}
In order to get the differential number flux of background neutrinos, 
first we compute the present number density of the background
neutrinos per unit energy from one GRB. The contribution of neutrinos 
emitted in the interval of the redshift $z \sim z+dz$ is given as,
\begin{equation}
dn_{\nu}^{\mr{ob}}(E_{\nu})=
R_{\mr{GRB}}(z){(1+z)}^{3}\frac{dt}{dz}dz 
\frac{dN_{\nu}(E_{\nu}^{'})}{dE_{\nu}^{'}}dE_{\nu}^{'}{(1+z)}^{-3},
\end{equation}
where 
\begin{equation}
\frac{dt}{dz}= - \frac{1}{H_0 (1+z)}\frac{1}{\sqrt{\Omega_{\Lambda}
+\Omega_{k}{(1+z)}^2+\Omega_{m}{(1+z)}^{3}}},
\end{equation}
and $E_{\nu}^{'}=(1+z)E_{\nu}$ is the energy of neutrinos at redshift
$z$, which is now observed as $E_{\nu}$ and 
$dN_{\nu}(E_{\nu}^{'})/dE_{\nu}^{'}$ is 
the number spectrum of neutrinos emitted by one GRB event. 
Hence, the GRB neutrino background can be calculated using the following
equation,
\begin{eqnarray}
\Phi_{\nu} \equiv \frac{cdn_{\nu}^{\mr{ob}}}{dE_{\nu}d\Omega} 
&=& \frac{c}{4\pi H_{0}} \int _{0}^{z_{\mr{max}}} dz \, 
R_{\mr{GRB}}(z) \frac{dN_{\nu}((1+z)E_{\nu})}
{dE_{\nu}^{'}}  \nonumber \\ &\times&
\frac{1}{\sqrt{\Omega_{\Lambda}+\Omega_{k}{(1+z)}^2+\Omega_{m}{(1+z)}^{3}}}.
\end{eqnarray}
In this paper, we adopt $z_{\mr{max}}=11$, $\Omega_{m}=0.3$,
$\Omega_{k}=0$, $\Omega_{\Lambda}=0.7$ and $H_0=71 \, \mr{km} 
\mr{s}^{-1} \mr{Mpc}^{-1}$.
Here, $R_{\mr{GRB}}(z)$ is the beaming-corrected (overall) GRB rate. The local
GRB rate has some uncertainties depending on the rate history, but
our results on the neutrino background are not so changed 
because the main contribution to the background comes from GRBs that 
occur at $z \sim (1-3)$, the number of which is observationally
determined. We use the following GRB rate history 
in units of ${\mr{yr}}^{-1}{\mr{Gpc}}^{-3}$ \cite{KM1,Gue3},
\begin{subequations}
\begin{eqnarray}
R_{\mr{GRB1}}(z) &=& 18  \frac{46 \, \mr{e}^{3.4z}}{\mr{e}^{3.8z}+45}
 F(z, \Omega_{m}, \Omega_{\Lambda}) \, \, \, \, \, \, \,  \, \, \, \, \, \, \\
R_{\mr{GRB2}}(z) &=& 18  \frac{23 \, \mr{e}^{3.4z}}{\mr{e}^{3.4z}+22} 
F(z, \Omega_{m}, \Omega_{\Lambda})  \, \, \, \, \, \, \, \, \, \, \, \, \, \,\\
R_{\mr{GRB3}}(z) &=& 23  \frac{24 \,
\mr{e}^{3.05z-0.4}}{\mr{e}^{2.93z}+15} F(z, \Omega_{m},
 \Omega_{\Lambda}) \, \, \, \, \, \, \, \, \, \, \, \, \, \, \, \, \, \, \\
R_{\rm{GRB4}}(z) &=& 43 F(z, \Omega_{m}, \Omega_{\Lambda}) 
\left\{\begin{array}{ll} {10}^{0.75z} \, (z
< 1) \\ {10}^{0.75} \, \, \, (z > 1) 
\end{array} \right. \, \, \, \, \,
\end{eqnarray}
\end{subequations}
where $F(z, \Omega_{m}, \Omega_{\Lambda}) = \sqrt{\Omega_{\Lambda}+ 
\Omega_m {(1+z)}^{3}}/{(1+z)}^{3/2}$.
In this paper, we show results calculated under the GRB3 model in most
figures, because the dependence of the resulting neutrino backgrounds
on GRB rate models is not so large \cite{KM1}. 
In the previous work \cite{KM1}, we evaluated the neutrino background by
exploiting Eqs. (B4) with the beaming-corrected energy $E_{\mr{GRB}} = 
1.24 \times {10}^{51}$ ergs. Alternatively, when we 
have isotropic energy, we should use the apparent GRB rate
which is expressed as $\rho_{\mr{GRB}} \equiv f_b R_{\mr{GRB}}$.
In this paper, we adopt $f_{b}=1/75$ as a fiducial value for
evaluation of the neutrino background from the RS. 

\section{\label{sec:level8}Vacuum Neutrino Oscillation}
Neutrino physics has been interesting since neutrino experiments recently
discovered something new, rather than giving only more precise
measurements of standard model parameters, or stronger bounds on
unseen new physics. Solar and atmospheric neutrino data directly show that
the flux of neutrinos with each flavor is not conserved, which
suggests that neutrinos are massive then there should be a neutrino
mixing matrix.

In GRBs, neutrinos are produced via photomeson production, so that 
we can expect $\nu_{e}:\nu_{\mu}:\nu_{\tau} \approx 1:2:0$ approximately. 
As a result of neutrino oscillation, we can obtain
$\nu_{e}:\nu_{\mu}:\nu_{\tau} \approx 1:1:1$. So there may be a 
possibility that tau neutrinos are detected through double bang 
events \cite{Ath1}.
However, we should note that the
magnetic field is strong in GRBs, so that contributions from 
muons that are more subject to cooling due to their longer life time,
are suppressed. Hence, we expect
$\nu_{e}:\nu_{\mu}:\nu_{\tau} \approx 0:1:0$ in the high energy region. As a
result of neutrino oscillation, we can obtain
$\nu_{e}:\nu_{\mu}:\nu_{\tau} \approx 1:1.8:1.8$ rather than 
$\nu_{e}:\nu_{\mu}:\nu_{\tau} \approx 1:1:1$ in the high energy region 
\cite{Kas1}. 

In this paper, we consider vacuum neutrino oscillation 
in the long baseline limit. Assuming that
$\theta_{23}$ is maximal ($\theta_{23} \approx \pi/4$) and
$\theta_{13}$ is very small ($\theta_{13} \approx 0$), which are
indicated by neutrino oscillation data, we can obtain,
\begin{eqnarray}
\Phi_{{\nu}_{e}} &\approx& \Phi_{{\nu}_{e}}^0 - \frac{1}{4} \sin^2 2
\theta_{12} (2 \Phi_{{\nu}_{e}}^0 - \Phi_{{\nu}_{\mu}}^0 - 
\Phi_{{\nu}_{\tau}}^0)\\
\Phi_{{\nu}_{\mu}} &\approx& \Phi_{{\nu}_{\tau}} \approx \frac{1}{2}
(\Phi_{{\nu}_{\mu}}^0 + \Phi_{{\nu}_{\tau}}^0) \nonumber \\ 
&\, &  \,\,\,\,\,\,\,\,\,\,\,\,\,\, + \frac{1}{8} \sin^2 2
\theta _{12} (2 \Phi_{{\nu}_{e}}^0 - \Phi_{{\nu}_{\mu}}^0 - 
\Phi_{{\nu}_{\tau}}^0).
\end{eqnarray}
In this paper, we adopt $\theta_{12}=0.59$.

\newpage 

\end{document}